\documentclass[letterpaper]{article} 
\usepackage{aaai2026}  
\usepackage{times}  
\usepackage{helvet}  
\usepackage{courier}  
\usepackage[hyphens]{url}  
\usepackage{graphicx} 
\usepackage{natbib}  
\usepackage{caption} 
\usepackage{algorithm}
\usepackage{algorithmic}

\usepackage{newfloat}
\usepackage{listings}
\DeclareCaptionStyle{ruled}{labelfont=normalfont,labelsep=colon,strut=off} 
\floatstyle{ruled}
\newfloat{listing}{tb}{lst}{}
\floatname{listing}{Listing}
\usepackage[utf8]{inputenc} 
\usepackage[T1]{fontenc}    
\usepackage{url}            
\usepackage{booktabs}       
\usepackage{amsfonts}       
\usepackage{nicefrac}       
\usepackage{microtype}      
\usepackage{xcolor}   
\usepackage{graphicx}
\usepackage{amsmath}
\usepackage{bm}
\usepackage{subfig}
\usepackage{algorithm}
\usepackage{algorithmic}
\usepackage{multirow}
\usepackage{makecell}
\usepackage{amssymb}

\newcommand{\methodname}{scCluBench}

\usepackage{pifont} 

\title{scCluBench: Comprehensive Benchmarking of
Clustering Algorithms for Single-Cell RNA Sequencing}
\author {
    Ping Xu\textsuperscript{\rm 1,3},
    Zaitian Wang\textsuperscript{\rm 1,3},
    Zhirui Wang\textsuperscript{\rm 2,3},
    Pengjiang Li\textsuperscript{\rm 1,3},
    Jiajia Wang\textsuperscript{\rm 1},
    Ran Zhang\textsuperscript{\rm 1,3},\\
    Pengfei Wang\textsuperscript{\rm 1,2,3,}
    \thanks{Corresponding author.},
    Yuanchun Zhou\textsuperscript{\rm 1,2,3}
}
\affiliations {
    \textsuperscript{\rm 1}Computer Network Information Center, Chinese Academy of Sciences, Beijing, China\\
    \textsuperscript{\rm 2}Hangzhou Institute for Advanced Study,University of Chinese Academy of Sciences, Hangzhou, China\\
    \textsuperscript{\rm 3}University of Chinese Academy of Sciences, Beijing, China\\
    xuping0098@gmail.com, wpf2106@gmail.com, zyc@cnic.cn
}

\usepackage{bibentry}

\begin{document}

\maketitle


\begin{abstract}

Cell clustering is crucial for uncovering cellular heterogeneity in single-cell RNA sequencing (scRNA-seq) data by identifying cell types and marker genes. 
Despite its importance, benchmarks for scRNA-seq clustering methods remain fragmented, often lacking standardized protocols and failing to incorporate recent advances in artificial intelligence. 
To fill these gaps, we present \textbf{scCluBench}, a comprehensive \textbf{bench}mark of \textbf{clu}stering algorithms for \textbf{sc}RNA-seq data.  
First, scCluBench provides 36 scRNA-seq datasets collected from diverse public sources, covering multiple tissues, which are uniformly processed and standardized to ensure consistency for systematic evaluation and downstream analyses.
To evaluate performance, we collect and reproduce a range of scRNA-seq clustering methods, including traditional, deep learning-based, graph-based, and biological foundation models. 
We comprehensively evaluate each method both quantitatively and qualitatively, using core performance metrics as well as visualization analyses. 
Furthermore, we construct representative downstream biological tasks, such as marker gene identification and cell type annotation, to further assess the practical utility. 
scCluBench then investigates the performance differences and applicability boundaries of various clustering models across diverse analytical tasks, systematically assessing their robustness and scalability in real-world scenarios. 
Overall,~\methodname~offers a standardized and user-friendly benchmark for scRNA-seq clustering, with curated datasets, unified evaluation protocols, and transparent analyses, facilitating informed method selection and providing valuable insights into model generalizability and application scope.\footnotemark[2] 

\footnotetext[2]{All datasets, code, and the Extended version for scCluBench are available at the link: 
\url{https://github.com/XPgogogo/scCluBench}.
More details for each stage are provided in the extended version.}

\end{abstract}

\section{Introduction}

Single-cell RNA sequencing (scRNA-seq) has transformed biological research by enabling the high-resolution exploration of cellular diversity, developmental processes, and tissue organization~\cite{shapiro2013single}. 
scRNA-seq clustering, which groups cells based on gene expression profiles, is a cornerstone analysis in scRNA-seq studies and underpins critical tasks such as cell type characterization, atlas construction, and marker gene discovery~\cite{kiselev2019challenges, wang2025sccompass}. 
As scRNA-seq datasets grow in size and complexity, the challenges of achieving robust, reproducible, and biologically meaningful clustering results become increasingly prominent, highlighting the urgent need for advanced computational techniques.
However, there is currently no comprehensive and standardized benchmarking framework for scRNA-seq clustering methods, making it difficult to objectively compare model performance, assess robustness and reproducibility across datasets, and select appropriate tools for specific biological contexts~\cite{xu2025scunified, krzak2019benchmark}. 

Powered by traditional and artificial intelligence methods, we propose \textbf{\methodname}, a comprehensive \textbf{bench}marking framework for \textbf{s}ingle-\textbf{c}ell RNA sequencing \textbf{clu}stering. 
~\methodname~ systematically compares clustering algorithms under unified conditions, providing standardized solutions in all major stages of scRNA-seq clustering benchmarking, including data resources, evaluation metrics, biological interpretation pipelines, and unified benchmarking workflows. 



\begin{figure*}[thbp]
    \centering
    \vspace{-3mm}
    \includegraphics[width=0.92\linewidth]{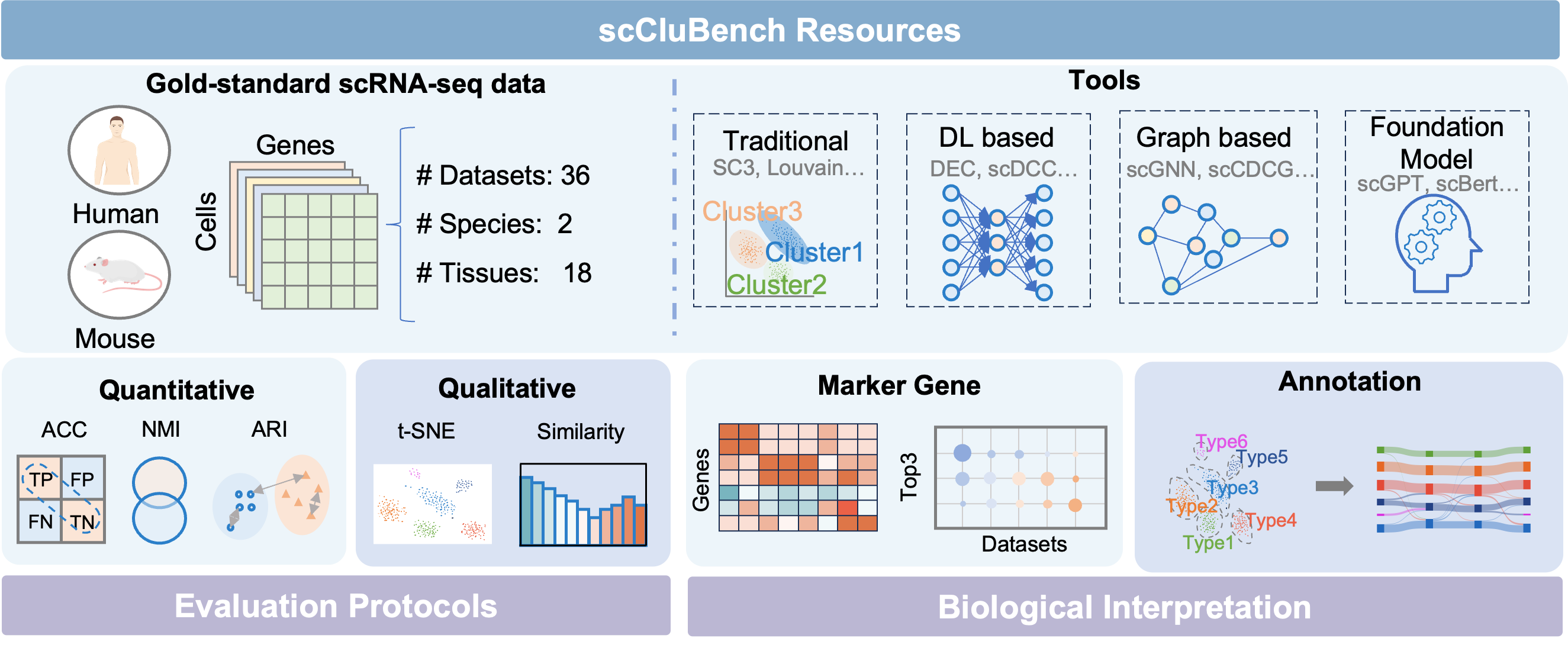}
    \vspace{-4mm}
    \caption{Overview of scCluBench: resources, evaluation protocols, and biological interpretation.}
    \label{fig: framework}
    \vspace{-3mm}
\end{figure*}

(1) \textbf{Standardization of benchmark resources.}
Existing scRNA-seq clustering benchmarks often lack dataset diversity, such as limited species or tissue types, and insufficient coverage of emerging models, particularly recent advances in biological foundation models built upon Transformer architectures.
\methodname~present a collection of 36 human and mouse datasets spanning diverse tissues.
This standardized resource, encompassing traditional, deep learning–based, graph-based, and foundation models, enables systematic evaluation and fair comparison of single-cell clustering methods.

(2) \textbf{Standardization of evaluation protocols.}
Assessment of scRNA-seq clustering methods often relies on limited quantitative and qualitative metrics. 
Thus, we standardize the evaluation process by incorporating diverse quantitative indicators on multiple datasets, along with qualitative assessments such as 2D visualization of cell embeddings. 
In particular, we offer quantitative analyses of embedding similarity-visualized to systematically evaluate phenomena like representation collapse and provide broader perspectives for model selection and optimization. 

(3) \textbf{Standardization of biological interpretation.}
Downstream analyses such as marker gene identification and cell type annotation are essential for interpreting clustering results, yet they are often inconsistently addressed. 
\methodname~deliver standardized, reproducible pipelines for marker gene detection and cell type labeling, complemented by gold-standard references for annotation. 
This ensures clustering outputs can be validated and interpreted in biological contexts, facilitating applications in single-cell research.

(4) \textbf{Unified benchmarking Workflow and Modular Code.} 
\methodname~provides an integrated and reproducible workflow covering data preprocessing, clustering, and cell type annotation. 
Standardized input–output formats and modularized implementations ensure ease of use and enable fair and consistent performance comparisons across all models.

By constructing~\methodname, we systematically enabled comparative analyses and identified several key findings: 

\begin{itemize}
    \item 
    We identified three critical components for fair and effective evaluation of scRNA-seq clustering methods: diverse and representative datasets, broad coverage computational methods, and a unified and reproducible analysis pipeline with standardized input/output formats. 

    \item 
    We find that existing scRNA-seq clustering methods suffer from distinct but significant limitations. 
    Traditional methods perform poorly in handling sparse, high-noise data. 
    Deep learning approaches, while effective in dimensionality reduction and denoising, often fail to capture underlying relationships between cells. 
    Graph-based models, although improving structural awareness, suffer from issues such as over-smoothing and embedding collapse.
    More fundamentally, most methods decouple embedding learning from clustering optimization, resulting in embedding spaces that are not fully conducive to clustering, thereby limiting overall performance.
    \item 
    We find that current scRNA-seq foundation models are often designed to construct a unified embedding space transferable to multiple downstream tasks, prioritizing general cell representation rather than task-specific optimization. Although such a general-purpose design enhances cross-task transferability, it also diminishes performance in specific tasks, such as clustering.
    
    \end{itemize}

\section{scCluBench}

\subsection{Benchmark Framework}
The~\methodname~framework offers an extensive benchmark for scRNA-seq clustering, as illustrated in Fig.~\ref{fig: framework}, which highlights its comprehensive dataset collection and incorporation of cutting-edge methods. 
It features a curated collection of 36 diverse datasets derived from human and mouse, spanning 18 tissue types, which serve as comprehensive testbeds for evaluating clustering algorithms. 
The benchmark encompasses a wide spectrum of scRNA-seq clustering methods, including traditional, deep learning-based, graph-based, and, notably, emerging biological foundation models. 
This diverse combination of datasets and clustering methods underpins a thorough evaluation, combining quantitative metrics and qualitative analyses, such as 2D cell embedding visualizations for comprehensive insights. 
Additionally,~\methodname~ standardizes biological interpretation through reproducible pipelines for marker gene detection and cell type annotation, ensuring clustering results are effectively validated within the context of real biological applications. 

\begin{figure*}[!t]
\centering

\subfloat[Number of Cells]{
\includegraphics[height=3cm,width=0.19\textwidth,keepaspectratio]{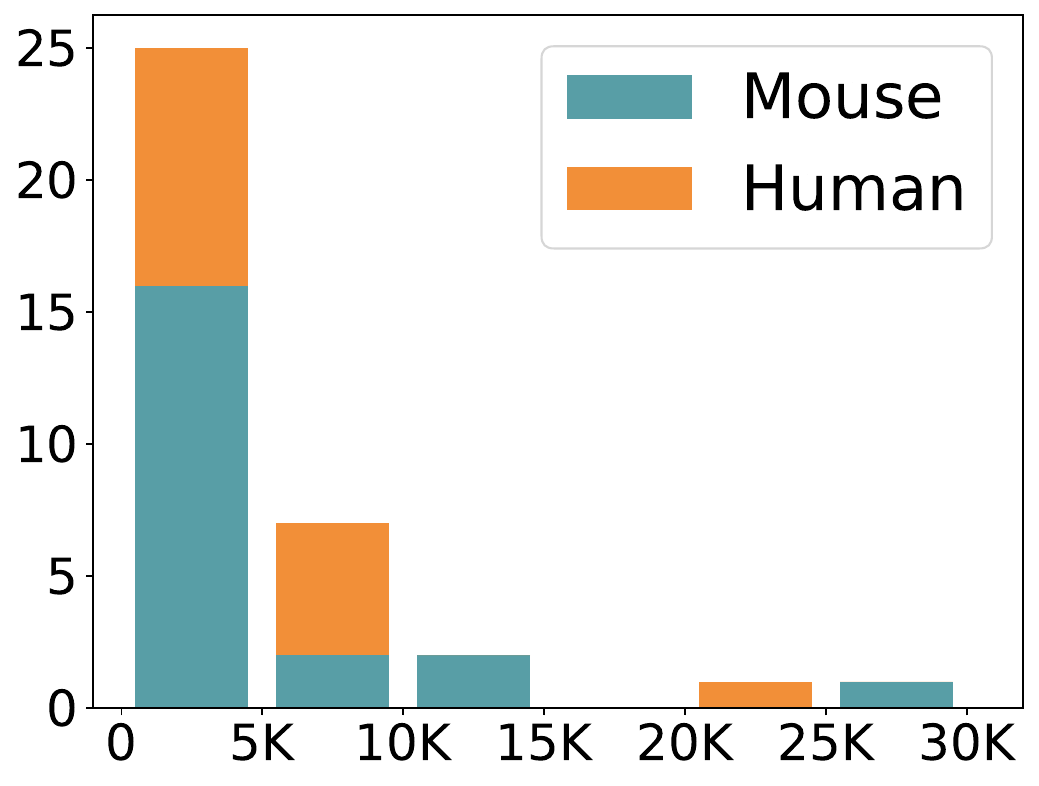}
\label{subfig:cell}
}
\subfloat[Number of Genes]{
\includegraphics[height=3cm,width=0.19\textwidth,keepaspectratio]{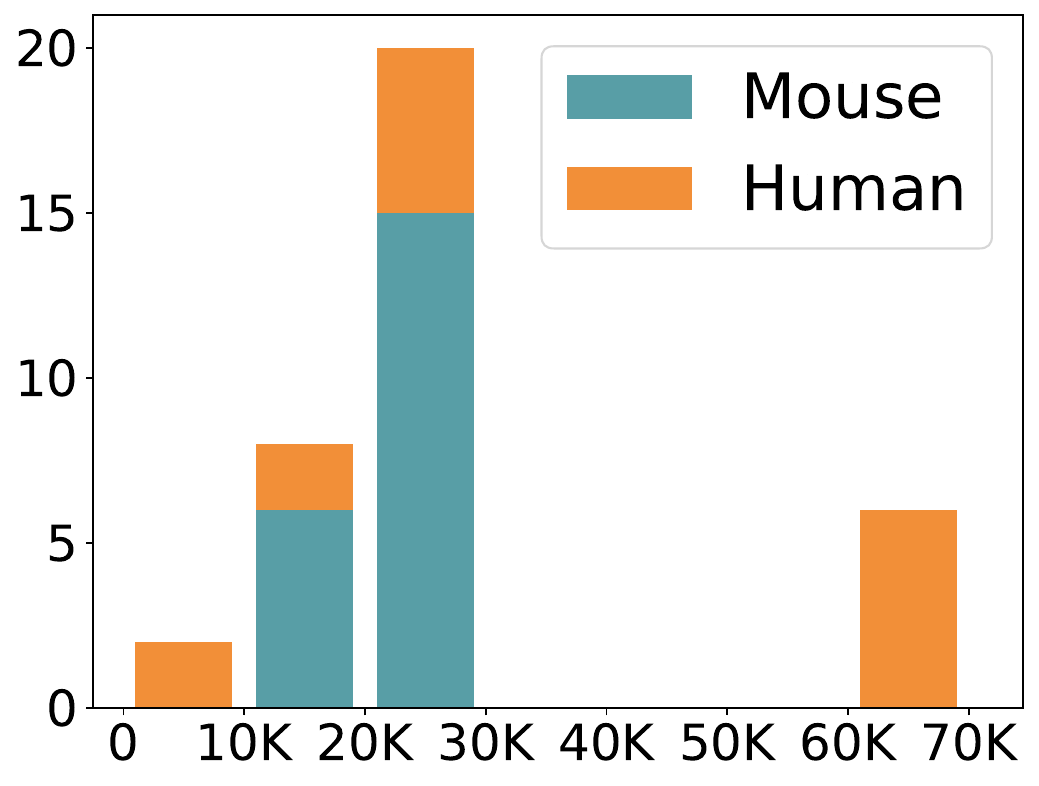}
\label{subfig:gene}
}
\subfloat[Number of Clusters]{
\includegraphics[height=3cm,width=0.19\textwidth,keepaspectratio]{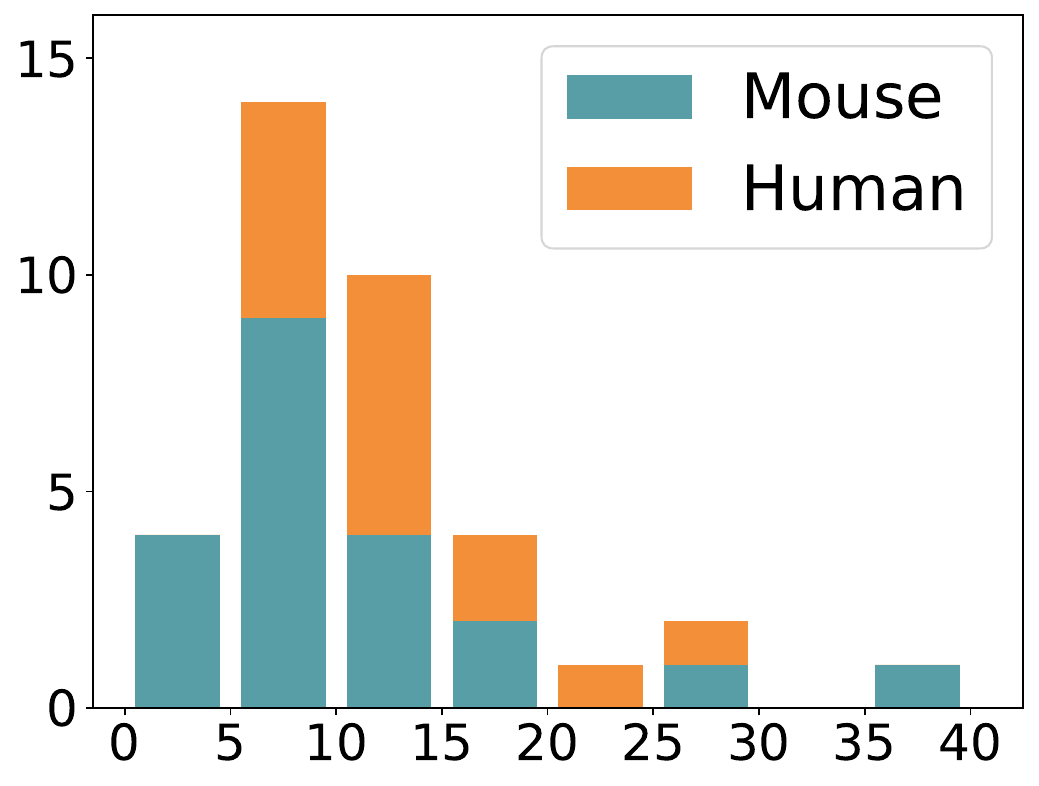}
\label{subfig:cluster}
}
\subfloat[Sparsity (\%)]{
\includegraphics[height=3cm,width=0.19\textwidth,keepaspectratio]{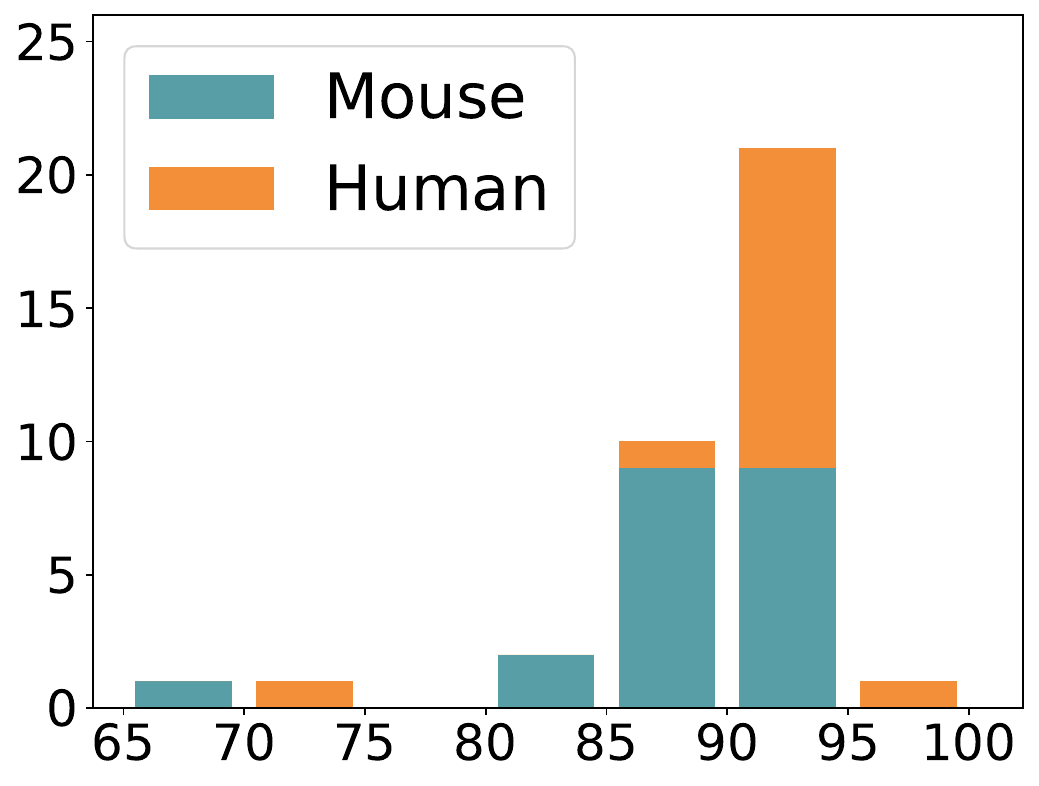}
\label{subfig:sparsity}
}
\\
\vspace{-4mm}
\subfloat[Distribution of Sample Numbers in Human and Mouse Data]{
\includegraphics[width=0.75\textwidth,keepaspectratio]{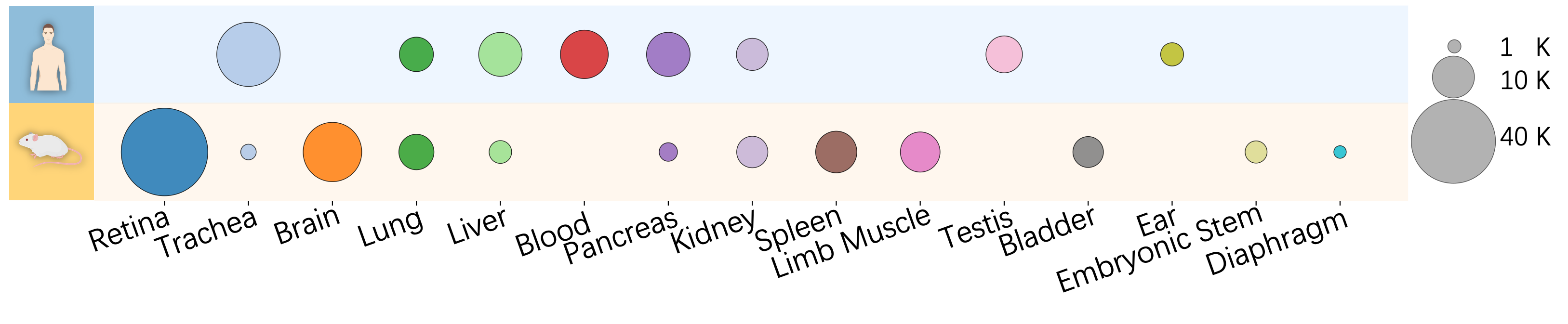}
\label{subfig:organ}
}
\vspace{-2mm}
\caption{Dataset distributions. (a) to (d) show dataset distributions by cell count, gene number, clusters, and sparsity, while (e) displays the distribution of samples between human and mouse datasets.
}
\label{fig:Data_Distribution}
\end{figure*} 

\begin{table*}[!t]  
 \centering  
 \tiny 
 \renewcommand\arraystretch{0.9}
 \resizebox{\textwidth}{!}{ 
 \begin{tabular}{ccccccc} 
 \toprule  
 \textbf{Methods} & \textbf{Methods} & \textbf{Framework} & \textbf{\# of Clusters} & \textbf{Language} & \textbf{Ref.} & \textbf{Journal}\\   
 \midrule  
 \multirow{3}*{\textbf{\makecell{Traditional \\ Models}}} 
 & SC3 & SingleCellExperiment & Automatic & R & \cite{kiselev2017sc3} & Nature Methods \\ 
  & Louvain & Seurat & Automatic & R & \cite{stuart2019comprehensive} & Cell \\ 
  & Leiden & Seurat & Automatic & R & \cite{stuart2019comprehensive} & Cell \\ 
 \midrule  
 \multirow{7}*{\textbf{\makecell{Deep Learning\\-based Models}}} 
 & DEC & AE & Automatic & Python & \cite{xie2016unsupervised} & ICML \\ 
  & scDeepCluster & AE & Hyperparameter & Python & \cite{tian2019clustering} & Nature Machine Intelligence \\ 
  & DESC & Stacked AE & Automatic & Python & \cite{li2020deep} & Nature Communications \\ 
  & scziDesk & AE & Hyperparameter & Python & \cite{chen2020deep} & NAR Genomics and Bioinformatics \\ 
  & scDCC & AE & Hyperparameter & Python & \cite{tian2021model} & Nature Communications \\ 
  & scNAME & Masked AE & Hyperparameter & Python & \cite{wan2022scname} & Bioinformatics \\ 
  & scMAE & Masked AE & Hyperparameter & Python & \cite{fang2024scmae} & Bioinformatics \\ 
 \midrule  
 \multirow{4}*{\textbf{\makecell{Graph-based \\ Models}}} 
 &scGNN & GNN & Hyperparameter & Python & \cite{wang2021scgnn} & Nature Communications \\ 
  &scDSC & AE + GNN & Hyperparameter & Python & \cite{gan2022deep} & Briefings in Bioinformatics \\ 
  &AttentionAE-sc & AE + GNN & Automatic & Python & \cite{li2023attention} & PLOS Computational Biology \\ 
  &scCDCG & Cut-informed graph embedding & Hyperparameter & Python & \cite{xu2024sccdcg} & DASFAA \\ 
 \midrule  
 \multirow{3}*{\textbf{\makecell{Foundation\\ Models}}} 
 &scGPT & Masked Language Model & Hyperparameter (Finetune) & Python & \cite{cui2024scgpt} & Nature Methods \\ 
  &GeneFormer & Context‑aware BERT & Hyperparameter (Finetune) & Python & \cite{theodoris2023transfer} & Nature \\ 
  & GeneCompass & Knowledge‑informed Transformer & Hyperparameter (Finetune) & Python & \cite{yang2024genecompass} & Cell Research \\ 
 \bottomrule 
 \end{tabular}  
 } 
\vspace{-3mm}
 \caption{Benchmark Clustering methods (AE: Autoencoder; GNN: Graph Neural Network).}  
 \vspace{-4mm}
 \label{tab:baseline} 
 \end{table*}

\subsection{Benchmark Datasets}
As data quality and diversity are critical to model performance~\cite{wang2025comprehensive,wang2025diversity}, 
~\methodname~comprises a diverse collection of 36 single-cell gene expression datasets from human and mouse specimens, covering 18 distinct tissue types, as shown in Fig.~\ref{fig:Data_Distribution}. 
Notably, scCluBench includes 2 large-scale datasets with over 20,000 cells and 5 high-dimensional datasets containing more than 60,000 genes. 
Additionally, 4 datasets contain at least 20 cell types, and 34 datasets exhibit sparsity rates exceeding 80\%, with overall sparsity levels ranging from 65.76\% to 95.42\%. 
Furthermore, these datasets are generated using diverse sequencing technologies, including CEL-seq, Drop-seq, Smart-seq2, and 10X Genomics, ensuring broad coverage of sequencing methods. 

\subsection{Benchmark Models} 
The~\methodname~benchmarks a representative collection of state-of-the-art (SOTA) clustering algorithms, spanning four methodological categories: traditional clustering, deep learning-based, graph-based, and biological foundation models, 
offering a diverse and standardized framework for systematic and fair evaluation of single-cell clustering performance.
A comprehensive list of all methods with brief descriptions is provided in Tab.~\ref{tab:baseline}.
All methods follow parameter settings from original publications. When unspecified, we perform controlled tuning to ensure stable and broadly applicable performance. 
Each dataset-method pair is independently run five times, 
and results are reported as mean ± standard deviation.

\subsection{Benchmark Evaluation}
\subsubsection{Evaluation Protocols.}
We propose a set of standardized evaluation protocols encompassing three quantitative metrics to reflect the correctness of cell clustering, and two qualitative assessment approaches to investigate the distinguishability of cell representations. 

\noindent\textbf{\textit{Quantitative Analysis.}}
The primary aim of single-cell clustering is to assign cells of each type with faithful and consistent class labels. 
Our evaluation of clustering performance focuses on three established metrics from the public domain: Accuracy (ACC), Normalized Mutual Information (NMI), and Adjusted Rand Index (ARI). 
ACC measures the optimal mapping between the ground-truth and predicted cluster algorithm assignments. 
NMI reflects the consistency between the predicted and true labels. 
ARI evaluates the similarity of the two assignments, ignoring permutations. 
For these measurements, higher values indicated better performance.

\noindent\textbf{\textit{Qualitative Analysis.}}
In machine learning, building accurate decision boundaries relies on distinguishable features. 
So it is important to understand the quality of cell representations learned by different methods and how the quality affects model performance.
To this end, we offer 2D visualizations of cell embeddings using the t-SNE dimensionality reduction technique to enable a more straightforward elaboration on how cell features are learned and clustered.
Additionally, we calculate representation similarities of all learned embeddings and qualitatively analyze the probability distributions. 
By observing embeddings' concentration on the high-similarity region, we can assess the severity of the over-smoothing problem. 
Through extensive qualitative analysis, our evaluation protocol provides better interpretability to models' performance and effectiveness. 

\subsubsection{Biological Interpretation.}
We propose a biological evaluation framework to systematically assess the concordance between predicted clusters and true cell types.

\noindent\textbf{\textit{Marker Gene Identification.}}
Differentially expressed genes (DEGs) are genes with significant expression differences across cell populations or experimental conditions.
Among these, genes that display strong cluster-specific expression patterns can be further selected as marker genes for cell type annotation.
We detected DEGs for each cluster with ``\texttt{rank\_genes\_groups}'' (default settings) from the Scanpy package. 
Using ground-truth cell type labels, 
we first extract the top 100 DEGs per reference cluster to form a \emph{gold-standard} marker list. The same procedure is then applied to clusters predicted by each model, yielding a comparable list of 100 marker genes per cluster.
We further plot the top 3 marker genes using the \textbf{tracksplot} diagram to provide straightforward presentations and compare how the expression values of DEGs stand out among other genes.

\noindent\textbf{\textit{Cell Type Annotation.}}
Cell type annotation serves as the primary downstream task and main objective of single-cell clustering. 
In our benchmark, we perform and compare two annotation approaches.
\textbf{a. Best-mapping annotation}: As a rapid approach, it directly aligns model-predicted cluster labels to ground-truth labels by maximizing one-to-one correspondence using the Hungarian algorithm, disregarding gene expression. 
\textbf{b. Marker-overlap annotation}: For each model-predicted cluster, we compute the proportion of shared genes between its marker list and the marker list of each gold-standard cluster. This overlap is calculated as \(\text{score}(p,g) = |\text{DEG}_p \cap \text{DEG}_g |/100\), where \(p\) and \(g\) are the model-predicted cluster label and gold-standard cluster label. The cluster's cell type is assigned to the gold-standard cell type with the highest overlap score. 
To highlight discrepancies between the two annotation methods and quantify their deviations from the gold standard labels, we construct \textbf{Sankey diagrams} to trace the relationships between best-mapping annotations, marker-overlap annotations, and gold standard cell types. 
We intend to use these visualization results to examine the efficacy and benefits of applying marker-based cell type annotation.

\begin{table*}[!th]
\centering
\resizebox{0.99\textwidth}{!}{
\setlength{\tabcolsep}{2pt}
\renewcommand{\arraystretch}{1}
\begin{tabular}{lcccccccccccccccc}
\toprule
\multicolumn{1}{c}{} &
\multicolumn{3}{c}{Traditional} & &
\multicolumn{7}{c}{Deep Learning-based} & &
\multicolumn{4}{c}{Graph-based} \\
\cmidrule{2-4} \cmidrule{6-12} \cmidrule{14-17}

& SC3 & louvain & leiden &
& DEC & DESC & scDeepCluster & scMAE & scNAME & scDCC & scziDesk &
& scGNN & scDSC & AttentionAE-sc & scCDCG \\

\midrule

Human Pancreas 1
 & 87.35±1.56 & 76.56±7.65 & 71.76±0.21 & 
 & 37.97±1.10 & 76.20±1.02 & 64.50±3.78 & \underline{87.99±3.41} & 75.94±12.61 & 69.51±3.25 & 68.13±1.47 & 
 & 56.65±0.42 & 58.57±5.15 & 81.13±1.88 & \textbf{92.15±0.84} \\
 
Human Pancreas 2
 & 83.87±1.49 & \textbf{90.66±0.69} & \underline{90.55±1.60} & 
 & 45.05±1.79 & 67.13±3.10 & 54.83±1.66 & 79.12±3.07 & 67.07±5.72 & 69.95±3.65 & 54.10±1.15 & 
 & 55.05±3.67 & 77.76±1.26 & 84.37±8.11 & 84.66±4.11 \\
 
Human Pancreas 3
 & 79.97±4.29 & 91.40±0.32 & \textbf{93.31±0.22} & 
 & 45.47±1.54 & \underline{91.93±3.99} & 50.88±3.15 & 90.11±3.94 & 71.01±5.37 & 63.62±7.48 & 78.78±12.26 & 
 & 67.84±4.53 & 83.73±0.59 & 91.20±2.79 & 88.04±0.34 \\
 
Human Pancreas4
 & 68.30±2.84 & 73.52±0.00 & 72.99±2.47 & 
 & 46.58±5.15 & 70.51±4.75 & 55.56±1.21 & 77.56±7.98 & 67.00±4.52 & 57.55±3.13 & 64.67±9.76 & 
 & 44.90±0.00 & 76.93±3.77 & \underline{82.33±4.65} & \textbf{86.57±0.29} \\
 
Mauro Pancreas
 & 83.51±1.77 & 88.69±0.00 & 92.08±0.13 & 
 & 65.61±1.10 & 76.22±2.33 & 73.55±1.37 & \textbf{95.62±0.16} & 89.73±9.92 & 86.72±9.05 & 69.97±17.08 & 
 & 64.84±2.25 & 70.31±5.59 & 92.43±2.72 & \underline{92.65±2.93} \\
 
68K PBMC
 & 79.44±0.23 & 65.04±0.09 & 60.88±6.50 & 
 & 59.36±1.12 & 55.48±3.54 & \underline{81.06±4.74} & 75.84±0.16 & 78.40±1.67 & \textbf{82.79±2.96} & 63.76±2.01 & 
 & 41.63±2.88 & 44.60±0.00 & 62.63±2.31 & 78.52±0.67 \\
 
CITE CMBC
 & \underline{73.67±2.99} & 70.51±18.30 & \textbf{78.04±3.27} & 
 & 38.67±5.51 & 67.07±2.29 & 65.95±2.27 & 72.91±5.55 & 63.53±3.14 & 72.44±2.94 & 47.47±12.26 & 
 & 64.56±2.49 & 30.78±0.59 & 64.77±8.75 & 71.45±1.85 \\
 
Human Kidney
 & 69.67±3.49 & 67.77±0.00 & 72.68±6.32 & 
 & 38.82±3.19 & 60.83±3.39 & 63.26±4.55 & \textbf{83.99±3.45} & 73.78±0.75 & 60.64±3.41 & \underline{80.08±0.57} & 
 & 40.44±0.00 & 40.09±0.00 & 63.77±4.54 & 79.55±0.29 \\
 
Sonya liver
 & 79.09±4.36 & 58.04±0.01 & 69.84±5.17 & 
 & 40.90±1.22 & 57.17±3.80 & 69.21±2.47 & \textbf{80.73±1.86} & \underline{79.97±8.80} & 70.64±4.59 & 76.58±1.52 & 
 & 33.60±0.00 & 42.44±0.09 & 78.60±8.70 & 75.34±3.67 \\
 
Sapiens Liver
 & \textbf{85.74±8.08} & \underline{79.30±0.00} & 71.19±0.00 & 
 & 65.76±3.20 & 42.24±0.00 & 49.08±1.90 & 67.51±2.30 & 63.33±1.70 & 57.88±4.60 & 68.19±0.60 & 
 & 73.83±3.00 & 64.67±4.40 & 68.76±11.20 & 73.09±1.40 \\
 
Sapiens Ear Crista Ampullaris
 & 56.96±10.69 & 44.84±0.00 & 43.37±0.94 & 
 & 57.73±2.07 & 29.83±0.00 & 53.73±2.50 & 67.00±0.40 & 67.35±0.70 & 80.67±3.70 & 39.58±0.00 & 
 & \underline{83.12±1.00} & 76.20±8.10 & 81.46±7.90 & \textbf{85.45±4.30} \\
 
Sapiens Ear Utricle
 & 59.62±0.00 & 53.07±0.00 & 51.21±0.00 & 
 & 59.38±8.40 & 60.23±0.00 & 54.83±1.20 & 73.16±0.60 & 71.95±1.40 & 68.51±5.70 & 73.58±0.60 & 
 & 69.89±2.30 & \textbf{84.94±6.70} & 62.49±9.30 & \underline{79.58±0.70} \\
 
Sapiens Lung
 & 51.75±0.00 & 53.58±1.44 & 48.38±0.11 & 
 & 60.11±5.00 & 55.50±0.00 & 44.07±1.70 & 63.24±1.10 & 59.09±2.30 & 57.40±3.20 & \underline{71.18±1.70} & 
 & \textbf{74.81±1.60} & 61.88±5.80 & 69.64±9.00 & 62.06±1.60 \\
 
Sapiens Testis
 & 42.63±0.00 & 62.93±13.22 & 62.86±0.00 & 
 & 45.38±1.50 & 20.79±0.00 & 35.17±1.70 & 53.71±0.80 & 63.05±9.10 & 43.89±1.00 & \underline{74.71±2.80} & 
 & \textbf{79.66±0.20} & 69.13±12.00 & 71.37±15.90 & 67.18±3.80 \\
 
Sapiens Trachea
 & 43.46±0.00 & 56.98±24.56 & 48.49±0.95 & 
 & 52.30±8.40 & 50.87±0.00 & 39.62±0.60 & 65.78±3.70 & 70.71±4.00 & 56.12±3.70 & 68.97±6.30 & 
 & 66.83±0.00 & \underline{71.45±6.00} & \textbf{87.85±4.93} & 52.46±2.90 \\
 
\midrule

Mouse cerebral cortex
 & \underline{80.53±0.40} & 65.32±0.47 & 73.64±0.00 & 
 & 49.87±9.58 & 58.62±3.53 & 73.60±2.73 & 72.99±0.62 & \textbf{80.70±1.59} & 74.27±6.53 & 76.68±3.65 & 
 & 36.22±0.22 & 32.80±0.62 & 71.80±2.36 & 71.73±0.12 \\
 
Mouse embryonic stem
 & 88.15±2.34 & 83.47±1.87 & 83.44±2.05 & 
 & 66.75±11.59 & 70.92±8.65 & \underline{97.47±0.42} & 80.73±1.93 & 85.38±0.52 & 73.86±7.73 & 88.82±1.61 & 
 & 62.84±5.00 & 71.84±0.00 & 78.77±2.90 & \textbf{98.96±0.06} \\
 
Mouse hypothalamus
 & 60.15±3.76 & 24.40±2.15 & 18.05±1.98 & 
 & 54.60±5.90 & 43.93±4.47 & 59.82±2.81 & \textbf{89.56±0.20} & 88.92±1.61 & 67.45±4.34 & \underline{89.54±0.39} & 
 & 38.25±0.00 & 38.49±0.55 & 79.13±6.91 & 85.31±0.34 \\
 
Mouse Pancreas 1
 & 58.15±2.67 & 73.84±1.92 & 73.36±2.41 & 
 & 48.25±1.92 & 62.14±5.63 & 43.45±1.51 & 74.18±7.57 & 60.49±3.43 & 51.39±4.84 & 68.76±3.08 & 
 & 47.93±4.35 & 49.05±2.96 & \underline{81.70±3.73} & \textbf{82.64±0.51} \\
 
Mouse Pancreass 2
 & 65.04±3.12 & 69.92±2.54 & 68.98±1.87 & 
 & 32.55±0.73 & 52.39±3.55 & 49.68±2.80 & \underline{91.89±0.20} & 72.62±8.95 & 57.38±0.34 & 74.81±12.83 & 
 & 42.54±2.91 & 75.47±2.83 & 84.53±7.39 & \textbf{93.97±0.23} \\
 
Shekhar mouse retina
 & 67.94±2.89 & 80.68±1.76 & 70.14±2.32 & 
 & 26.17±2.84 & 86.47±8.72 & 63.83±4.61 & \textbf{93.51±0.02} & \underline{89.93±0.43} & 70.21±1.77 & 51.72±0.25 & 
 & 27.99±0.74 & 37.10±6.26 & \textcolor{gray}{OOM} & 76.04±1.85 \\
 
Macosko mouse retina 
 & 64.88±2.45 & 70.04±1.98 & 63.58±2.17 & 
 & 31.60±2.37 & \underline{84.85±2.20} & 54.52±2.13 & \textbf{87.68±1.01} & 80.15±5.16 & 62.74±3.71 & 72.65±3.59 & 
 & 27.15±0.00 & 42.96±0.00 & \textcolor{gray}{OOM} & 69.78±0.70 \\
 
Mouse Kidney
 & \underline{92.27±1.87} & 69.32±2.34 & 81.07±1.65 & 
 & 24.65±1.83 & 68.07±0.00 & 75.60±2.13 & \textbf{93.45±0.17} & 87.49±7.45 & 80.59±5.39 & 89.14±9.02 & 
 & 20.70±0.13 & 19.89±3.22 & 30.33±16.14 & 61.47±1.07 \\
 
Mouse bladder
 & 63.29±2.76 & 73.81±1.92 & \textbf{82.52±1.43} & 
 & 50.12±3.02 & \underline{76.09±5.14} & 62.94±3.62 & 66.21±6.92 & 64.94±2.70 & 68.99±0.97 & 43.09±6.75 & 
 & 52.32±3.35 & 46.83±1.55 & 52.27±3.90 & 75.61±1.23 \\
 
QS Diaphragm
 & 94.48±1.23 & 78.97±2.45 & 78.51±1.87 & 
 & 49.96±7.16 & 65.06±9.82 & 71.34±0.35 & \textbf{98.97±0.11} & 98.01±0.29 & 39.77±7.58 & 71.38±15.54 & 
 & 50.25±2.99 & 56.71±2.85 & 95.83±1.21 & \underline{98.71±0.25} \\
 
QS Lung
 & 53.70±2.87 & 66.11±1.76 & 52.68±2.34 & 
 & 36.99±1.27 & 57.24±2.13 & 47.93±3.00 & 74.50±1.21 & 69.77±1.02 & 51.19±10.54 & \underline{76.31±4.18} & 
 & 41.86±0.84 & 49.29±2.23 & \textbf{76.73±3.92} & 70.58±0.87 \\
 
QS Trachea
 & 80.96±2.13 & 39.04±3.21 & 57.26±2.87 & 
 & 47.51±6.05 & 33.90±3.74 & 67.41±0.90 & 82.62±10.50 & 82.86±5.37 & 59.14±3.47 & \textbf{85.88±5.56} & 
 & 48.07±2.86 & 49.48±16.76 & 79.10±10.18 & \underline{85.42±0.03} \\
 
QS Limb Muscle
 & 91.65±1.45 & 73.58±2.76 & 71.28±2.34 & 
 & 49.05±0.43 & 52.46±5.65 & 69.30±8.48 & \textbf{98.96±0.14} & \underline{98.35±0.24} & \textcolor{gray}{ERR} & 89.66±7.11 & 
 & 47.28±0.89 & 50.25±0.58 & 87.33±6.52 & 92.94±1.00 \\
 
Qx Limb Muscle
 & 83.09±2.34 & 58.79±3.12 & 76.08±1.98 & 
 & 75.82±0.81 & 51.15±8.49 & 79.35±4.08 & \textbf{99.05±0.86} & \underline{98.73±0.45} & 84.54±3.37 & 97.25±0.87 & 
 & 56.15±0.36 & 61.73±0.00 & 97.47±1.23 & 96.51±0.60 \\
 
Qx Bladder
 & 77.40±2.56 & 46.52±3.45 & 48.28±2.87 & 
 & 74.56±4.61 & 52.38±2.56 & 78.53±0.84 & 84.41±12.91 & 91.96±12.92 & 73.58±3.63 & \textbf{99.59±0.09} & 
 & 79.12±1.32 & 77.32±6.28 & 94.80±3.81 & \underline{98.84±0.31} \\
 
Qx Spleen
 & 55.79±2.87 & 43.63±3.21 & 43.81±2.65 & 
 & 48.84±6.22 & 53.81±8.00 & 65.28±0.88 & \underline{96.06±0.18} & \textbf{96.52±1.66} & 65.91±16.76 & 96.04±1.24 & 
 & 59.46±1.97 & 75.60±3.65 & 87.47±13.77 & 94.48±0.49 \\
 
Muris Limb Muscle
 & \textbf{98.60±0.87} & \underline{97.13±1.23} & 96.72±1.45 & 
 & 54.79±6.50 & 39.22±0.00 & 59.57±3.90 & 66.13±3.40 & 61.34±3.10 & 70.38±4.20 & 53.31±4.30 & 
 & 48.62±2.30 & 64.37±4.00 & 53.35±10.50 & 94.50±7.10 \\
 
Muris Brain
 & 54.60±3.21 & 33.90±2.87 & 40.84±3.45 & 
 & 55.70±3.20 & 15.02±0.00 & 85.36±18.10 & 71.37±0.00 & 90.24±0.30 & 65.02±2.00 & \textcolor{gray}{OOM} & 
 & 91.40±0.10 & \textbf{96.02±2.50} & 73.41±26.09 & \underline{95.55±1.10} \\
 
Muris Kidney
 & \underline{65.29±2.34} & 44.10±3.12 & 38.16±2.76 & 
 & 47.46±2.60 & 49.42±7.54 & 42.30±5.20 & 55.52±3.40 & 47.47±2.30 & 56.94±5.20 & 41.97±3.10 & 
 & 46.48±1.50 & 36.38±2.80 & 46.32±10.60 & \textbf{80.65±1.60} \\
 
Muris Liver
 & 48.86±2.98 & \underline{55.86±2.45} & 45.72±3.21 & 
 & 46.51±5.10 & 48.90±0.00 & 42.62±3.20 & 53.48±0.40 & 49.72±4.10 & 45.39±4.30 & 44.50±3.40 & 
 & 51.58±2.90 & 55.76±7.50 & 41.04±8.70 & \textbf{68.13±1.40} \\
 
Muris Lung
 & 42.54±3.12 & 40.35±2.87 & 50.45±2.65 & 
 & 50.54±3.80 & 53.26±0.00 & 37.98±2.40 & 51.06±2.20 & 38.15±1.80 & 50.10±1.80 & 37.73±4.30 & 
 & 40.98±4.80 & 36.85±4.20 & \underline{64.54±26.60} & \textbf{65.68±1.70} \\

\midrule

AVG
& 70.34±3.56 & 64.49±4.21 & 65.06±2.87 & 
 & 49.48±3.83 & 57.15±3.28 & 60.64±3.07 & 78.24±2.60 & 74.88±3.69 & 64.78±4.51 & 69.96±4.47 & 
 & 53.75±1.77 & 57.71±3.65 & 74.08±7.48 & 81.29±1.45 \\

\midrule

Model Rank AVG
& 5 & 9 & 7 &
& 14 & 12 & 10 & 2 & 3 & 8 & 6 &
& 13 & 11 & 4 & 1 \\
  
\bottomrule
\end{tabular}
}
\vspace{-2mm}
\caption{ACC scores (mean ± std) across 36 datasets; the best score is shown in \textbf{bold}, and the second-best is \underline{underlined}.}
\vspace{-2mm}
\label{tab:ACC_label}
\end{table*}

\begin{table*}[!th]
  \centering
  \tiny
  \renewcommand\arraystretch{0.3}
  \setlength{\tabcolsep}{1.6pt}
  {
    \renewcommand{\arraystretch}{1}
    \begin{tabular}{lcccccccccccccccccc}
      \toprule
      \multirow{5}{*}{Dataset} 
      & \multicolumn{8}{c}{Clustering Performance} &
      & \multicolumn{8}{c}{Classification Performance} \\
      \cmidrule{2-9} \cmidrule{11-18}
      
      & \multicolumn{2}{c}{scGPT} &
      & \multicolumn{2}{c}{GeneFormer} &
      & \multicolumn{2}{c}{GeneCompass} &
      & \multicolumn{2}{c}{scGPT} &
      & \multicolumn{2}{c}{GeneFormer} &
      & \multicolumn{2}{c}{GeneCompass} \\
      \cmidrule{2-3} \cmidrule{5-6} \cmidrule{8-9} 
      \cmidrule{11-12} \cmidrule{14-15} \cmidrule{17-18}
      
      & ACC & F1 &
      & ACC & F1 &
      & ACC & F1 &
      & ACC & F1 &
      & ACC & F1 &
      & ACC & F1 \\
      
      \midrule

Sapiens Ear Crista Ampullaris
  & {\scriptsize 52.28}±2.37 & {\scriptsize 45.78}±2.70 &
  & {\scriptsize 20.57}±0.47 & {\scriptsize 13.82}±0.13 &
  & {\scriptsize 41.24}±1.91 & {\scriptsize 27.28}±1.56 &
  & {\scriptsize 98.14}±1.06 & {\scriptsize 95.77}±2.84 &
  & {\scriptsize 93.39}±0.98 & {\scriptsize 84.64}±4.30 &
  & {\scriptsize 94.92}±1.04 & {\scriptsize 88.50}±2.46 \\

Sapiens Ear Utricle
  & {\scriptsize 51.33}±0.87 & {\scriptsize 45.34}±3.77 &
  & {\scriptsize 31.29}±2.00 & {\scriptsize 26.43}±2.53 &
  & {\scriptsize 36.56}±0.55 & {\scriptsize 26.53}±1.71 &
  & {\scriptsize 97.10}±3.10 & {\scriptsize 96.72}±2.72 &
  & {\scriptsize 82.90}±0.88 & {\scriptsize 63.49}±5.88 &
  & {\scriptsize 98.06}±1.77 & {\scriptsize 96.26}±3.36 \\

Sapiens Liver
  & {\scriptsize 43.47}±4.37 & {\scriptsize 35.17}±2.47 &
  & {\scriptsize 24.49}±1.10 & {\scriptsize 22.14}±2.31 &
  & {\scriptsize 27.77}±1.46 & {\scriptsize 20.88}±0.75 &
  & {\scriptsize 88.43}±3.09 & {\scriptsize 72.72}±6.43 &
  & {\scriptsize 80.00}±1.44 & {\scriptsize 54.56}±4.32 &
  & {\scriptsize 87.78}±1.21 & {\scriptsize 70.96}±1.40 \\

Sapiens Lung
  & {\scriptsize 41.39}±1.85 & {\scriptsize 37.56}±2.12 &
  & {\scriptsize 12.54}±0.43 & {\scriptsize 10.35}±0.41 &
  & {\scriptsize 24.75}±1.21 & {\scriptsize 13.88}±0.65 &
  & {\scriptsize 93.38}±1.91 & {\scriptsize 84.39}±3.70 &
  & {\scriptsize 83.52}±0.37 & {\scriptsize 63.74}±3.88 &
  & {\scriptsize 87.44}±1.61 & {\scriptsize 76.60}±2.12 \\

Sapiens Testis
  & {\scriptsize 50.99}±3.03 & {\scriptsize 44.32}±6.01 &
  & {\scriptsize 18.74}±0.64 & {\scriptsize 8.76}±0.20 &
  & {\scriptsize 43.45}±7.35 & {\scriptsize 17.55}±1.41 &
  & {\scriptsize 97.52}±1.22 & {\scriptsize 92.11}±6.84 &
  & {\scriptsize 96.51}±0.58 & {\scriptsize 75.90}±4.78 &
  & {\scriptsize 96.90}±0.76 & {\scriptsize 84.58}±9.14 \\

Sapiens Trachea
  & {\scriptsize 39.14}±0.73 & {\scriptsize 32.85}±2.23 &
  & {\scriptsize 8.66}±0.31 & {\scriptsize 4.98}±0.07 &
  & {\scriptsize 18.46}±2.23 & {\scriptsize 9.76}±0.97 &
  & {\scriptsize 97.96}±0.39 & {\scriptsize 84.95}±1.78 &
  & {\scriptsize 96.37}±0.00 & {\scriptsize 77.39}±0.56 &
  & {\scriptsize 97.92}±0.00 & {\scriptsize 88.93}±0.00 \\
  
\midrule

Muris Brain
  & {\scriptsize 59.54}±0.64 & {\scriptsize 39.49}±0.28 &
  & {\scriptsize 62.71}±0.11 & {\scriptsize 40.64}±0.04 &
  & {\scriptsize 54.82}±0.02 & {\scriptsize 37.25}±0.01 &
  & {\scriptsize 99.84}±0.10 & {\scriptsize 98.01}±1.58 &
  & {\scriptsize 99.67}±0.14 & {\scriptsize 95.88}±1.49 &
  & {\scriptsize 100.00}±0.00 & {\scriptsize 100.00}±0.00 \\

Muris Kidney
  & {\scriptsize 61.92}±5.31 & {\scriptsize 51.54}±6.24 &
  & {\scriptsize 29.87}±1.37 & {\scriptsize 23.79}±1.15 &
  & {\scriptsize 18.70}±0.04 & {\scriptsize 13.32}±0.65 &
  & {\scriptsize 96.59}±1.52 & {\scriptsize 96.58}±2.42 &
  & {\scriptsize 77.25}±3.86 & {\scriptsize 74.21}±3.78 &
  & {\scriptsize 93.85}±3.83 & {\scriptsize 93.04}±3.44 \\

Muris Limb Muscle
  & {\scriptsize 29.22}±0.28 & {\scriptsize 21.88}±1.42 &
  & {\scriptsize 23.25}±1.29 & {\scriptsize 17.22}±0.43 &
  & {\scriptsize 24.47}±0.05 & {\scriptsize 19.76}±0.09 &
  & {\scriptsize 97.05}±1.25 & {\scriptsize 94.89}±1.94 &
  & {\scriptsize 90.41}±1.00 & {\scriptsize 80.38}±2.20 &
  & {\scriptsize 96.63}±0.76 & {\scriptsize 94.53}±1.30 \\

Muris Liver
  & {\scriptsize 32.44}±2.46 & {\scriptsize 20.99}±1.30 &
  & {\scriptsize 13.84}±0.50 & {\scriptsize 9.26}±0.29 &
  & {\scriptsize 28.80}±2.73 & {\scriptsize 19.42}±1.38 &
  & {\scriptsize 95.52}±1.22 & {\scriptsize 89.87}±3.57 &
  & {\scriptsize 86.71}±1.50 & {\scriptsize 59.18}±4.27 &
  & {\scriptsize 97.55}±0.00 & {\scriptsize 94.91}±0.00 \\

Muris Lung
  & {\scriptsize 14.28}±0.48 & {\scriptsize 13.29}±0.71 &
  & {\scriptsize 8.34}±0.22 & {\scriptsize 5.95}±0.13 &
  & {\scriptsize 12.87}±0.46 & {\scriptsize 10.66}±0.64 &
  & {\scriptsize 94.82}±1.36 & {\scriptsize 89.64}±2.92 &
  & {\scriptsize 80.23}±4.16 & {\scriptsize 54.68}±8.04 &
  & {\scriptsize 94.58}±0.00 & {\scriptsize 84.53}±0.00 \\
\bottomrule
\end{tabular}
}
\vspace{-2mm}
\caption{Clustering and classification performance(means ± std over 5 runs) of biological foundation models.}
\vspace{-5mm}
\label{tab:LLM-based Performance}
\end{table*}

\section{Observation and Analysis}
\subsection{Quantitative Analysis}
\subsubsection{Overall Clustering Performance.}
Tab.~\ref{tab:ACC_label} summarizes the clustering accuracy of each method on single-cell clustering tasks, where scCDCG stands out among all methods owing to its cut-informed graph embedding mechanism. 
By integrating detailed dataset characteristics, we derive the following consistent observations:
(1) \textit{Traditional methods} perform well on datasets with fewer than 5,000 cells and moderate gene dimensions. 
However, as the scale and complexity of data increase, their reliance on low-dimensional distance metrics leads to a marked decline in accuracy and stability.
(2) \textit{Among deep learning-based methods}, scMAE demonstrates robustness across varying data scales and sparsity levels. Its self-supervised feature reconstruction effectively mitigates expression bias caused by sparsity. 
scNAME ranks second, exhibiting certain robustness but performing slightly inferior on large-scale datasets.
(3) \textit{Graph-based methods} show distinct advantages in handling sparse data; however, their performance varies depending on graph construction strategies. 
Compared to hard graphs based on binary adjacency relationships, soft graphs, exemplified by the continuous edge-weight mechanism in scCDCG, provide a more refined characterization of intercellular similarities and differences, resulting in improved clustering accuracy on complex datasets. 
In summary, although traditional methods offer simplicity and interpretability in simple tasks, deep learning and graph neural network methods, with superior modeling capabilities and robustness, are more favorable for large-scale, high-sparsity single-cell data analysis.




\subsubsection{Performance of Biological Foundation Models.}
To evaluate biological foundation models for single-cell data, we conducted classification and clustering experiments, with results summarized in Tab.~\ref{tab:LLM-based Performance}. 
Compared with the clustering methods listed in Tab.~\ref{tab:ACC_label}, biological foundation models exhibit a marked performance gap in clustering tasks. 
Specifically, GeneFormer consistently underperforms in clustering accuracy across most datasets, whereas scGPT shows moderate gains on select datasets.
Conversely, these models achieve consistently higher accuracy and F1 score in classification tasks, with stable performance across multiple datasets, indicating superior generalization capabilities.
This performance gap reveals a limitation of current foundation models, which prioritize learning generalizable cell representations to enable broad transferability across downstream tasks but often sacrifice task-specific optimization, especially for fundamental tasks such as clustering that require dedicated mechanisms.

\subsubsection{Program Exceptions.}

During our experiments, we have observed some exceptions. 
Some out-of-memory errors occur when handling large datasets, such as when processing \textit{Macosko mouse retina} (14K+ samples) and \textit{Shekhar mouse retina} (20K+ samples) with AttentionAE-sc and processing \textit{Muris Brain} (13K+ samples) with scziDesk. 
In a rare case, namely when running scDCC on \textit{QS Limb Muscle}, the program fails due to a NaN-valued loss error.

\subsection{Qualitative Analysis}
\subsubsection{Cell Cluster Visualization.}
\begin{figure*}[!t]
    \centering
    \vspace{-3mm}
    \includegraphics[width=0.95\linewidth]{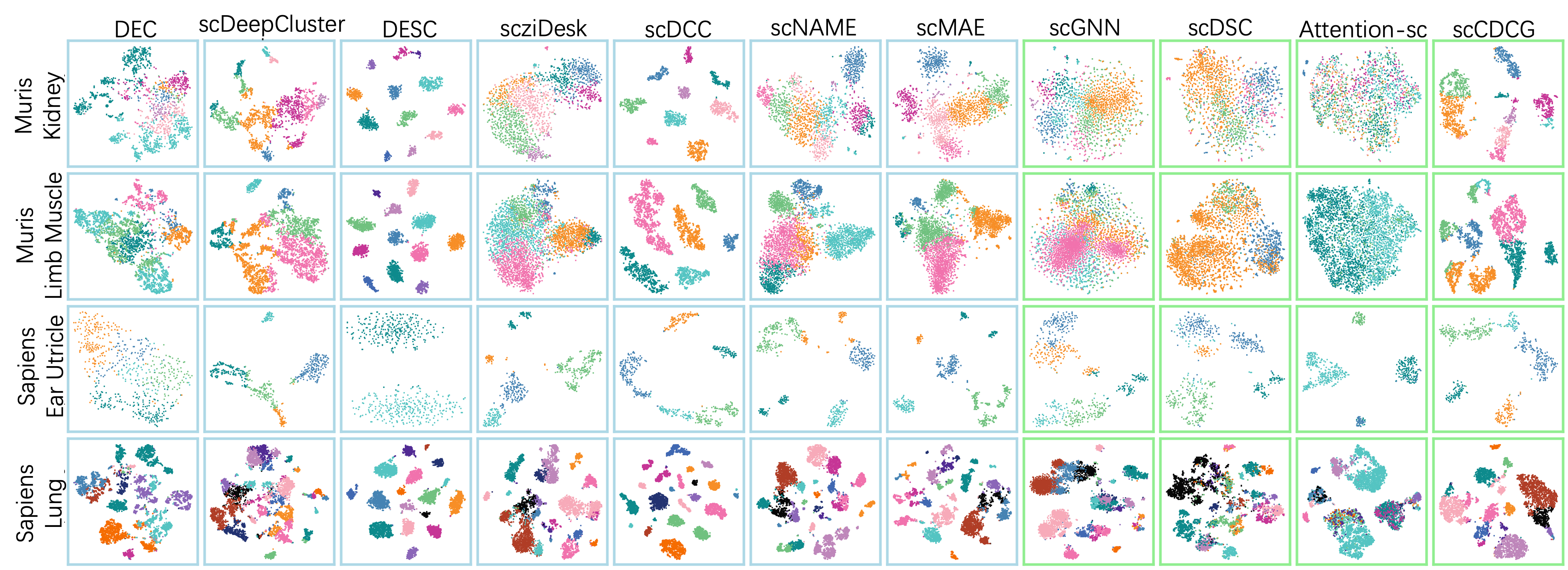}
    \caption{A digest of the visualization of all baselines.}
    \label{fig:vis_digest}
\end{figure*}

To enhance the interpretability of the quantitative results, we perform dimensionality reduction on the embeddings derived by each method and plot the cells on a 2D plane (Fig.~\ref{fig:vis_digest}).
We highlight 3 considerations to evaluate the quality of cell clustering through the visualization: (1) the number of clusters, (2) the boundary of each cluster, and (3) the compactness of cells within each cluster.
Regarding the number of clusters, we notice that some methods frequently generate mismatched assignments with ground-truth labels. 
For example, DESC assigns cells to 12 clusters for \textit{Muris Limb Muscle}, while the ground-truth has 6 cell types; it assigns cells to 2 clusters for \textit{Sapiens Ear Utricle}, while the ground-truth has 5 cell types. Such inconsistency leads to inferior clustering accuracies as shown in Tab.~\ref{tab:ACC_label}. 
As for the boundary, we notice that some graph-based methods yield vague boundaries, such as those by scDSC and scGNN, likely related to their inferior performance compared with scCDCG, which boasts a much clearer boundary and better performance. 
In terms of compactness, we can see that scDCC represents cells of the same categories with highly similar embeddings and locates them in proximal locations, suggesting that common patterns of each type are well-recognized, in accordance with its decent performance.

\subsubsection{Cell Representation Distinguishability.}
\begin{figure*}[!t]
    \centering
    \vspace{-4mm}
    \includegraphics[width=0.95\linewidth]{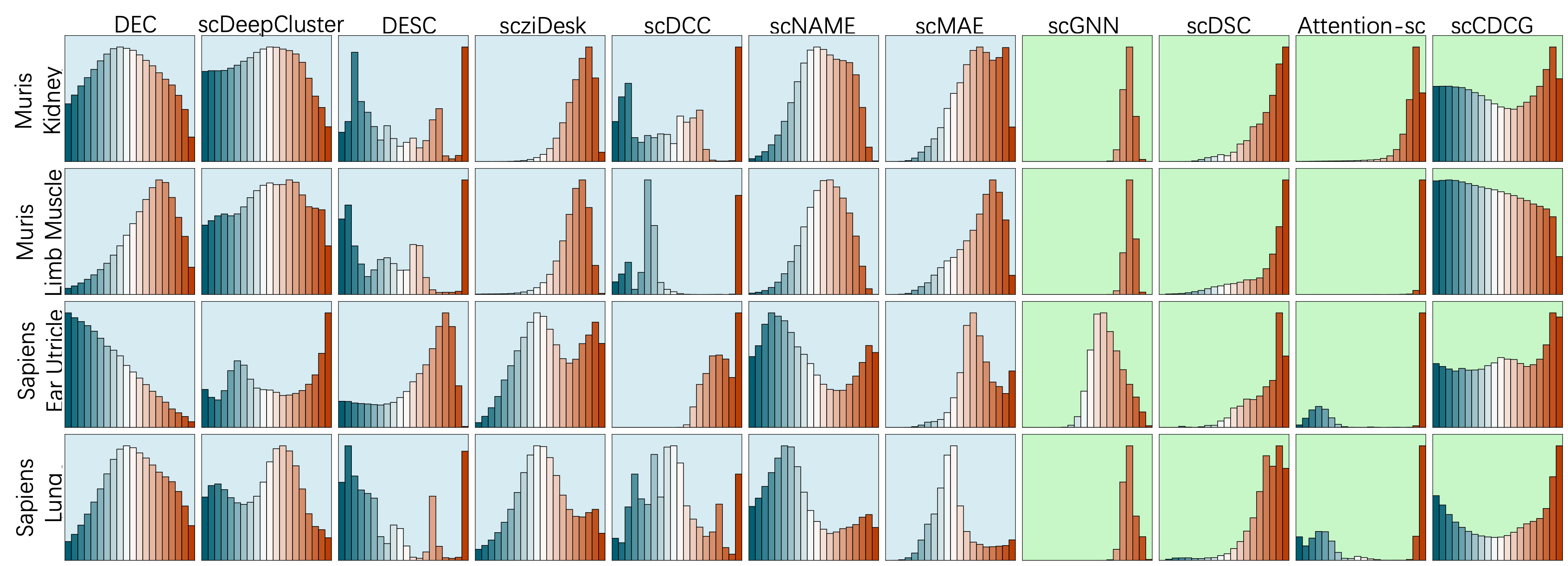}
    \vspace{-2mm}
    \caption{A digest of representation similarity of all baselines.}
    \vspace{-5mm}
    \label{fig:bar_digest}
\end{figure*}

Despite effectively leveraging graph structures, graph-based clustering methods, particularly GNN-based ones, often suffer from over-smoothing and representation collapse~\cite{wang2024deep,ning2025deep,ning2025rethinking,ning2021lightcake,ning2022graph}, where node representations across classes become indistinguishable,  due to the inductive bias of the GNN models that adjacent nodes are highly similar. 
To investigate the extent of the over-smoothing problem of each method, we calculate pair-wise cosine similarities for all sample embeddings in each dataset and derive their probability distribution. 
A representative digest of embedding similarity distribution is illustrated in Fig.~\ref{fig:bar_digest}, where red bars indicate the probabilities of embedding pairs with high similarity (up to 1), and blue bars indicate the probabilities of embedding pairs with low similarity (down to 0). 
By observing the similarity distribution of all methods, we can discover that deep learning methods are free from over-smoothing and representation collapse problems, with a substantial portion of embeddings considerably dissimilar and easily distinguished, as suggested by the blue and white bars. 
Graph-based methods, namely, AttentionAE-sc, scGAE, scDSC, and scGNN, on the other hand, suffer from severe representation collapse, where almost all embeddings are highly similar and indistinguishable. 
The only exception is scCDCG, which, thanks to its cut-informed graph embedding mechanism, outperforms other graph-based methods and even some deep learning approaches in embedding distinguishability.

\subsection{Biological Analysis}

\subsubsection{Case Study.}
\begin{figure*}[!t]
    \centering
\vspace{-3mm}
\includegraphics[width=1\linewidth]{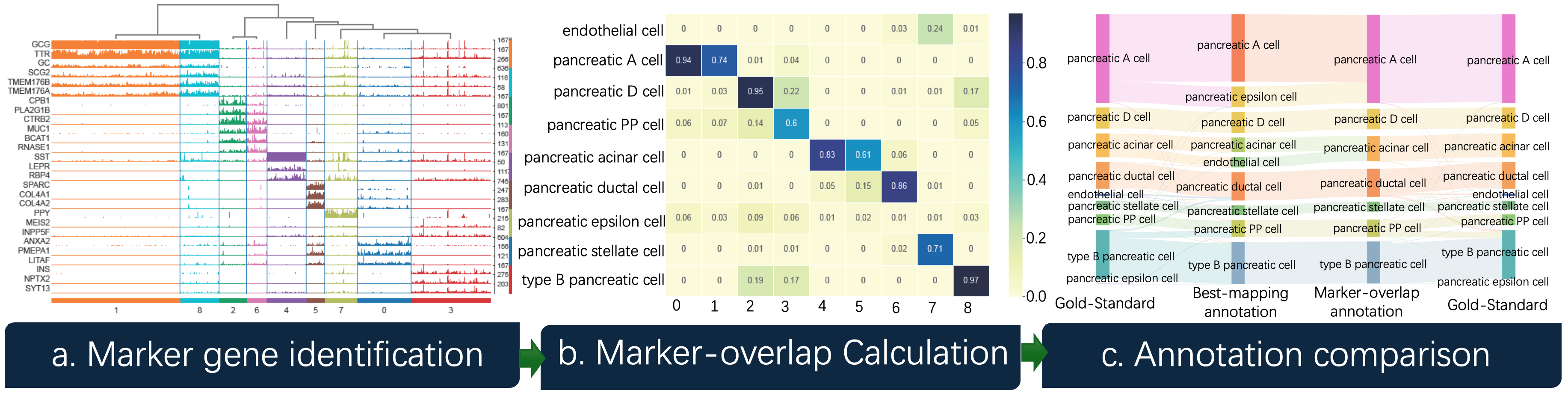}
\vspace{-5mm}
    \caption{All biological analysis of scCDCG on \textit{Mauro Human Pancreas cells}.}
    \label{fig:bio_scCDCG}
\end{figure*}

Taking the results of the scCDCG model on \textit{Mauro Pancreas} as an example (Fig.~\ref{fig:bio_scCDCG}), we performed an extensive biological analysis, including marker gene identification, cell type annotation tasks, and annotation comparison.

\noindent\textbf{\textit{Marker Gene Identification.}}
To evaluate marker gene expression across different clusters, we employed tracksplot to visualize the expression patterns of the top three marker genes predicted by the scCDCG model (Fig.~\ref{fig:bio_scCDCG}a). For instance, GCG, TTR, and GC were identified as the marker genes for cluster 1, whereas SCG2, TMEM76B, and TMEM76A were assigned to cluster 8. 
Tracksplot provides a clear visualization of gene expression across cell clusters, enabling the evaluation and refinement of cluster label assignments. 
Notably, clusters 1 and 8 exhibited similar expression profiles among their highly expressed genes, 
suggesting that these clusters may represent subtypes within a broader cell category, rather than entirely distinct cell populations. 
The top 3 marker genes may be insufficient to distinguish these two clusters.


\noindent\textbf{\textit{Cell Type Annotation.}}
To evaluate the biological interpretability of the clustering results, we performed cell type annotation on scCDCG-predicted clusters using marker-overlap annotation method (detailed in Evaluation Section). 
As shown in Fig.~\ref{fig:bio_scCDCG}b, clusters 3 and 1 were annotated as ``type B pancreatic cell'' and ``pancreatic A cell'', respectively.
Notably, ``endothelial cell'' and ``pancreatic epsilon cell'' were severely underrepresented (3 and 21 samples, respectively) in the reference dataset.
Indeed, the scarcity of reference samples increases the annotation difficulty for rare cell types.
Nevertheless, scCDCG successfully annotated the remaining seven major cell types with high accuracy.


\begin{table*}[!t]
\centering
\tiny
\renewcommand\arraystretch{0.7}
\setlength{\tabcolsep}{3.2pt}
{ 
\begin{tabular}{lcccccccccccccccccccccc}
\toprule
\multirow{2}{*}{Dataset} & \multicolumn{2}{c}{DEC} & \multicolumn{2}{c}{DESC} & \multicolumn{2}{c}{scDeepCluster} & \multicolumn{2}{c}{scNAME} & \multicolumn{2}{c}{scMAE} & \multicolumn{2}{c}{scDCC} & \multicolumn{2}{c}{scziDesk} & \multicolumn{2}{c}{scGNN} & \multicolumn{2}{c}{scDSC} & \multicolumn{2}{c}{AttentionAE-sc} & \multicolumn{2}{c}{scCDCG} \\
\cmidrule(lr){2-3} \cmidrule(lr){4-5} \cmidrule(lr){6-7} \cmidrule(lr){8-9} \cmidrule(lr){10-11} \cmidrule(lr){12-13} \cmidrule(lr){14-15} \cmidrule(lr){16-17} \cmidrule(lr){18-19} \cmidrule(lr){20-21} \cmidrule(lr){22-23}
 & BM & MO & BM & MO & BM & MO & BM & MO & BM & MO & BM & MO & BM & MO & BM & MO & BM & MO & BM & MO & BM & MO \\
\midrule
Mauro Pancreas & 65.27 & 83.74 & 78.98 & 95.29 & 75.87 & 94.63 & 22.15 & 21.21 & 95.52 & 95.52 & 74.65 & 85.34 & 85.30 & 90.29 & 68.24 & 77.66 & 50.71 & 54.85 & 81.90 & 87.23 & 80.44 & 93.50 \\
Sonya Liver & 50.83 & 47.23 & 60.45 & 88.70 & 55.77 & 89.38 & 21.57 & 15.60 & 86.96 & 95.67 & 64.93 & 84.18 & 65.73 & 79.41 & 42.25 & 63.23 & 32.77 & 38.11 & 72.06 & 77.57 & 48.95 & 73.61 \\
Sapiens Ear Utricle & 65.96 & 65.96 & 60.23 & 60.23 & 53.19 & 62.36 & 71.03 & 88.22 & 73.81 & 91.49 & 70.54 & 86.74 & 74.30 & 90.18 & 69.56 & 84.29 & 90.18 & 90.18 & 73.32 & 93.94 & 74.30 & 90.02 \\
Sapiens Liver & 62.59 & 66.45 & 42.24 & 64.31 & 51.72 & 73.05 & 61.52 & 68.40 & 68.49 & 70.77 & 53.35 & 68.12 & 67.61 & 71.24 & 66.64 & 65.94 & 66.08 & 66.91 & 85.73 & 80.34 & 50.42 & 67.29 \\
Sapiens Lung & 62.54 & 67.24 & 55.50 & 70.32 & 45.39 & 54.29 & 62.34 & 76.19 & 61.67 & 83.71 & 55.60 & 76.00 & 71.26 & 79.62 & 68.39 & 82.40 & 62.79 & 72.57 & 64.84 & 64.84 & 63.92 & 67.17 \\
Sapiens Testis & 43.98 & 77.92 & 20.8 & 75.1 & 38.40 & 65.44 & 54.00 & 79.92 & 54.84 & 80.88 & 44.52 & 80.20 & 80.29 & 79.86 & 79.81 & 84.89 & 66.93 & 84.39 & 99.29 & 100.00 & 71.67 & 80.08 \\
Sapiens Trachea & 54.59 & 78.23 & 50.86 & 83.18 & 38.99 & 65.57 & 68.78 & 90.75 & 64.08 & 89.89 & 58.87 & 85.3 & 64.27 & 85.84 & 56.67 & 84.04 & 80.17 & 80.22 & 90.41 & 90.47 & 51.15 & 72.96 \\
Muris Brain & 50.73 & 50.73 & 86.15 & 80.76 & 58.96 & 58.96 & 84.65 & 97.86 & 71.37 & 97.86 & 68.82 & 68.82 & - & - & 91.54 & 91.54 & 92.70 & 97.86 & 100.00 & 100.00 & 94.61 & 94.61 \\
Muris Kindey & 45.35 & 38.86 & 51.79 & 70.89 & 36.10 & 36.43 & 47.66 & 51.73 & 54.38 & 51.51 & 48.43 & 61.36 & 43.64 & 43.81 & 41.11 & 44.63 & 41.66 & 40.51 & 94.88 & 94.88 & 57.62 & 56.58 \\
Muris Limb Muscle & 49.29 & 62.54 & 39.22 & 92.53 & 55.56 & 66.02 & 61.82 & 84.07 & 71.18 & 71.85 & 67.70 & 81.32 & 51.34 & 76.42 & 43.66 & 60.10 & 67.60 & 67.68 & 29.11 & 76.65 & 60.03 & 82.98 \\
Muris Liver & 42.11 & 66.07 & 50.65 & 87.58 & 48.79 & 71.70 & 38.75 & 41.48 & 52.75 & 80.59 & 42.85 & 74.40 & 47.71 & 60.16 & 56.17 & 71.49 & 56.10 & 68.24 & 97.48 & 97.48 & 53.13 & 79.61 \\
Muris Lung & 52.78 & 75.58 & 51.36 & 79.99 & 37.47 & 66.00 & 47.42 & 42.54 & 48.29 & 73.74 & 48.13 & 77.49 & 40.97 & 0.54 & 34.88 & 54.13 & 31.82 & 38.11 & 50.07 & 50.07 & 39.09 & 55.24 \\
\midrule
ACC Mean & 54.23 & 65.31 & 52.09 & 80.19 & 50.21 & 67.76 & 54.62 & 65.35 & 66.94 & 82.70 & 60.15 & 78.5 & 62.26 & 72.85 & 61.78 & 73.39 & 63.29 & 68.53 & 79.20 & 85.65 & 63.43 & 76.68 \\

\midrule

Mean gain & \multicolumn{2}{c}{11.08} & \multicolumn{2}{c}{28.1} & \multicolumn{2}{c}{17.55} & \multicolumn{2}{c}{10.73} & \multicolumn{2}{c}{15.76} & \multicolumn{2}{c}{18.35} & \multicolumn{2}{c}{10.59} & \multicolumn{2}{c}{11.61} & \multicolumn{2}{c}{5.24} & \multicolumn{2}{c}{6.45} & \multicolumn{2}{c}{13.25} \\
\bottomrule
\end{tabular}
}
\vspace{-2mm}
\caption{Accuracy correction performance (BM: Best-mapping; MO: Marker-overlap).}
\vspace{-5mm}
\label{tab:result_correction}
\end{table*}

\noindent\textbf{\textit{Annotation Comparison.}}
Further, we employed the best-mapping annotation method to assign cell types to the clusters predicted by scCDCG. Fig.~\ref{fig:bio_scCDCG}c illustrates the correspondence among the best-mapping annotation, marker-overlap annotation, and the gold standard cell types, highlighting the differences between the two annotation strategies and quantifying their deviations from the ground truth. 
While the best-mapping method preserves the number of clusters consistent with reference labels, it often introduces ambiguous assignments (e.g., ``pancreatic A cell'' simultaneously annotated as both ``pancreatic A cell'' and ``pancreatic epsilon cell'', or erroneous merging of distinct types). 
In contrast, the marker-overlap annotation effectively rectifies these errors, and its corrections to the best-mapping annotations are also reflected in the figure.
Nevertheless, this method also faces limitations with extremely small populations (e.g., 3 ``endothelial cells'', 21 ``pancreatic epsilon cells''), where insufficient marker gene expression impedes accurate annotation.
The results indicate that, compared to the best-mapping method, the marker-overlap annotation approach provides greater biological interpretability by aligning more closely with gene expression patterns and established biological knowledge.

\subsubsection{Result Correction.}
The results of the two annotation methods were compared against the gold standard cell types, and ACC was calculated to more accurately reflect the model’s performance in cell type identification. 
During the evaluation, the ACC obtained by the best-mapping annotation was consistent with the average values reported in Tab.~\ref{tab:ACC_label} across five experiments, with differences falling within the expected statistical variance. 
Notably, the marker-overlap annotation effectively corrected specific misclassifications, leading to performance improvements across all methods. 
The revised results are summarized in Tab.~\ref{tab:result_correction}.
Overall, across the majority of datasets, the marker-overlap annotation achieved higher ACC than the best-mapping annotation, indicating that the incorporation of biological prior knowledge not only improves the accuracy of clustering performance evaluation but also enhances the biological interpretability of the clustering results.


\section{Related work}

Clustering methods for scRNA-seq have evolved from traditional models grounded in low-dimensional distance metrics to techniques leveraging deep learning and graph-based modeling, and most recently, to biological foundation models built upon Transformer architectures. 
Early approaches such as SC3, Louvain, and Leiden are limited by low-dimensional assumptions, restricting their ability to capture complex cellular heterogeneity. 
Deep learning-based methods (e.g., scMAE, scDeepCluster) enhance robustness against data sparsity and noise through unsupervised feature reconstruction, yet often suffer from instability and limited interpretability. 
Graph-based approaches, exemplified by scSiameseClu~\cite{xu2025scsiameseclu} and scSGC~\cite{xu2025soft},  further improve clustering accuracy by incorporating intercellular relationships, though they remain sensitive to graph construction strategies.
Recently, biological foundation models like scGPT and GeneCompass have achieved broad generalization through large-scale pretraining, yet their clustering performance remains limited by non-specific task design. 

Although several studies have benchmarked scRNA-seq clustering methods across aspects such as parameter sensitivity, cell number estimation, batch effect correction, and spatial transcriptomics, most evaluations remain limited to specific methodological categories or assessment dimensions~\cite{yuan2024benchmarking, dai2022scimc, yu2022benchmarking, tran2020benchmark}. 
For instance,~\cite{krzak2019benchmark} provided an early systematic evaluation of scRNA-seq clustering, but focused solely on R-based algorithms. 
To date, a comprehensive benchmarking framework spanning the full spectrum of clustering approaches, from traditional models to biological foundation models, has yet to be established. 
A unified platform integrating diverse clustering algorithms and evaluation metrics is essential for systematic benchmarking and further methodological advancement in single-cell analysis.

\section{Conclusion}

We present \textbf{\methodname}, a comprehensive and standardized benchmarking framework for scRNA-seq clustering that integrates diverse datasets, algorithmic paradigms, and multidimensional evaluation protocols.
The~\methodname~systematically compares traditional, deep learning, graph-based, and foundation model, offering detailed insights into their performance trade-offs and applicability boundaries across diverse clustering scenarios, thereby informing future method development and practical tool selection.
Looking ahead,~\methodname~will be expanded to include larger-scale datasets, integrate multi-modal single-cell data, and refine benchmarking protocols to address emerging biological challenges.
\clearpage
\newpage
\appendix
\section{Supplementary Msaterials}
\section{Data Details}
\subsection{Dataset Coverage}

\label{app:appendix_dataset}
scCluBench comprises a comprehensive collection of 36 single-cell RNA-seq gene expression datasets derived from human and mouse specimens, encompassing 18 distinct tissue types. These datasets exhibit substantial diversity in terms of data scale, dimensionality, cellular composition, and sequencing protocols. Specifically, the benchmark includes 2 large-scale datasets with over 20,000 cells and 5 high-dimensional datasets containing more than 60,000 genes. Additionally, 4 datasets feature at least 20 annotated cell types, reflecting complex cellular heterogeneity, while 34 datasets exhibit gene expression sparsity rates exceeding 80\%, with overall sparsity ranging from 65.76\% to 95.42\%.

The datasets span a wide range of sequencing technologies, including CEL-seq, Drop-seq, Smart-seq2, 10X Genomics, and Microwell-seq, covering diverse experimental platforms and biological contexts. This rich and varied collection presents both opportunities and challenges for clustering algorithm evaluation, ensuring a rigorous assessment of method robustness, scalability, and generalizability.

Comprehensive dataset details, including species, organ type, cell counts, gene numbers, cluster annotations, sparsity, and sequencing methods, are provided in Table~\ref{tab:dataset}.
\begin{table*}[!thbp]
    \centering
    \scriptsize
    \vspace{-5mm}
    \setlength{\tabcolsep}{4pt}
    \renewcommand{\arraystretch}{1}
    \begin{tabular}{llccccccc}
    \toprule
    \textbf{Species} & \textbf{Dataset Name} & \textbf{\#Cell} & \textbf{\#Gene} & \textbf{\#Cluster} & \textbf{Organ} & \textbf{Seq. Method} & \textbf{Sparsity (\%)} & \textbf{Ref.} \\
    \midrule
    \multirow{15}{*}{\textbf{\makecell{Human}}} 
    & Human Pancreas 1 & 1,937 & 20,125 & 14 & Pancreas & CEL-seq & 90.44 & \cite{baron2016single} \\
    & Human Pancreas 2 & 1,724 & 20,125 & 14 & Pancreas & CEL-seq & 90.59 & \cite{baron2016single} \\
    & Human Pancreas 3 & 3,605 & 20,125 & 14 & Pancreas & CEL-seq & 91.30 & \cite{baron2016single} \\
    & Human Pancreas 4 & 1,303 & 20,125 & 14 & Pancreas & CEL-seq & 89.05 & \cite{baron2016single} \\
    & Mauro Pancreas & 2,122 & 19,046 & 9 & Pancreas & CEL-seq2 & 73.02 & \cite{muraro2016single} \\
    & 68K PBMC & 4,271 & 16,653 & 8 & Blood & 10X Genomics & 92.24 & \cite{zheng2017massively} \\
    & CITE CMBC & 8,617 & 2,000 & 15 & Blood & 10X Genomics & 93.26 & \cite{zheng2017massively} \\
    & Human kidney  & 5,685 & 25,125 & 11 & Kidney & 10X Genomics & 92.92 & \cite{young2018single} \\
    & Sonya Liver & 8,444 & 4,999 & 11 & Liver & 10X Genomics & 90.77 & \cite{macparland2018single} \\
    & Sapiens Liver & 2,152 & \textbf{61,759} & 15 & Liver & Smart-seq2 & 95.42 & \cite{tabula2020single} \\
    & Sapiens Ear Crista Ampullaris & 2,357 & \textbf{61,759} & 7 & Ear & Smart-seq2 & 93.59 & \cite{tabula2020single} \\
    & Sapiens Ear Utricle & 611 & \textbf{61,759} & 5 & Ear & Smart-seq2 & 93.75 & \cite{tabula2020single} \\
    & Sapiens Lung & 6,530 & \textbf{61,759} & 25 & Lung & Smart-seq2 & 93.88 & \cite{tabula2020single} \\
    & Sapiens Testis & 7,494 & \textbf{61,759} & 8 & Testis & Smart-seq2 & 93.91 & \cite{tabula2020single} \\
    & Sapiens Trachea & \textbf{22,592} & \textbf{61,759} & 20 & Trachea & Smart-seq2 & 94.73 & \cite{tabula2020single} \\
    \midrule
    \multirow{22}{*}{\textbf{\makecell{Mouse}  }} 
    & Mouse cerebral cortex & 3,005 & 19,972 & 9 & Cerebral Cortex & Fluidigm C1 & 81.21 & \cite{zeisel2015cell} \\
    & Mouse embryonic stem & 2,717 & 24,175 & 4 & Embryonic Stem & Droplet Barcoding & 65.76 & \cite{klein2015droplet} \\
    & Mouse hypothalamus & 2,881 & 24,341 & 7 & Hypothalamus & CEL-seq & 87.77 & \cite{romanov2017molecular} \\  
    & Mouse Pancreas 1 & 822 & 14,878 & 13 & Pancreas & CEL-seq & 90.48 & \cite{baron2016single} \\
    & Mouse Pancreas 2 & 1,064 & 14,878 & 13 & Pancreas & CEL-seq & 89.80 & \cite{baron2016single} \\
    & Shekhar mouse retina & \textbf{27,499} & 13,166 & 19 & Retina & Drop-seq & 93.33 & \cite{shekhar2016comprehensive} \\
    & Macosko mouse retina & 14,653 & 11,422 & 39 & Retina & Drop-seq & 88.34 & \cite{macosko2015highly} \\  
    & Mouse Kidney & 3,660 & 23,797 & 8 & Kidney & Drop-seq & 92.32 & \cite{adam2017psychrophilic} \\
    & Mouse bladder & 2,746 & 20,670 & 16 & Bladder & Microwell-seq & 94.87 & \cite{han2018mapping} \\
    & QS Diaphragm & 870 & 23,341 & 5 & Diaphragm & Smart-seq2 & 91.35 & \cite{schaum2018single} \\
    & QS Lung & 1,767 & 23,341 & 11 & Lung & Smart-seq2 & 89.08 & \cite{schaum2018single} \\
    & QS Trachea & 1,350 & 23,341 & 4 & Trachea & Smart-seq2 & 85.48 & \cite{schaum2018single} \\
    & QS Limb Muscle & 1,090 & 23,341 & 6 & Limb Muscle & Smart-seq2 & 89.47 & \cite{schaum2018single} \\
    & Qx Limb Muscle & 3,909 & 13,341 & 6 & Limb Muscle & 10X Genomics & 83.57 & \cite{schaum2018single} \\
    & Qx Bladder & 2,500 & 23,341 & 4 & Bladder & 10X Genomics & 86.93 & \cite{schaum2018single} \\
    & Qx Spleen & 9,552 & 23,341 & 5 & Spleen & 10X Genomics & 94.34 & \cite{schaum2018single} \\
    & Muris Limb Muscle & 3,855 & 21,609 & 6 & Limb Muscle & Smart-seq2 & 91.38 & \cite{tabula2020single} \\
    & Muris Brain & 13,417 & 21,609 & 2 & Brain & Smart-seq2 & 91.83 & \cite{tabula2020single} \\
    & Muris Kidney & 1,817 & 21,609 & 9 & Kidney & Smart-seq2 & 92.25 & \cite{tabula2020single} \\
    & Muris Liver & 2,859 & 21,609 & 11 & Liver & Smart-seq2 & 88.20 & \cite{tabula2020single} \\
    & Muris Lung & 5,167 & 21,609 & 25 & Lung & Smart-seq2 & 89.90 & \cite{tabula2020single} \\
    \bottomrule
    \end{tabular}
    \caption{Details of 36 single-cell gene expression datasets across multiple sequencing platforms. Large datasets (with \( > 20000\) cells) and high-dimensional datasets (with \( > 60000\) genes) are highlighted with bold fonts. } 
    \label{tab:dataset}
\end{table*}

\subsection{Dataset Format}
All scRNA-seq datasets~\methodname~are provided in the \texttt{.h5ad} format, which ensures compatibility with standard single-cell analysis frameworks such as Scanpy and supports efficient data storage and manipulation.
Datasets can be loaded using \texttt{scanpy.read\_h5ad("path/to/file.h5ad")}, which returns an AnnData object.

Each dataset contains the following key components:
\begin{itemize}
    \item \textbf{Gene names:} 
    
    stored in \texttt{data.var["feature\_name"]}, representing standardized gene identifiers.
    \item \textbf{Cell type annotations:} 
    
    recorded in \texttt{data.obs["cell\_type"]}, providing ground-truth labels for each cell.
    \item \textbf{Expression matrix:} 
    
    accessible via \texttt{data.to\_df()}, returning a cell-by-gene matrix in Pandas DataFrame format, where rows represent cells and columns represent genes.
\end{itemize}

This standardized data structure facilitates consistent extraction of expression profiles and cell-type labels, enabling reproducible and scalable benchmarking across diverse datasets.

\subsection{Data Preprocessing}
In this study, we implemented a rigorous and standardized preprocessing pipeline to ensure consistency and comparability across single-cell RNA sequencing datasets. 
Raw data were loaded from \texttt{.h5ad} files into AnnData objects, containing gene expression matrices and corresponding cell annotations. 
We systematically assessed whether the data had undergone normalization, log-transformation (log1p), and scaling. 
For datasets lacking normalization, library size normalization was performed to mitigate technical variability associated with sequencing depth. 
Cell-specific size factors were then computed to standardize expression levels across cells. 
Untransformed data were subjected to log1p transformation to address the inherent skewness in expression distributions. 
Finally, z-score scaling was applied to center and scale gene expression values, ensuring uniform feature variance. 
This comprehensive preprocessing framework not only guarantees data quality and stability but also provides a standardized input foundation that facilitates fair and robust evaluation of downstream clustering algorithms.

\section{Models Details}
\label{app:appendix_models}

\subsection{Traditional Clustering Models}

\noindent$\bullet$ SC3 is an unsupervised clustering method for single-cell RNA-seq data that integrates multiple clustering solutions through consensus clustering, enhancing robustness and accuracy while effectively addressing data sparsity and noise~\cite{kiselev2017sc3}.

\noindent$\bullet$ The Louvain algorithm is a community detection method that optimizes modularity to identify clusters by maximizing the density of connections within clusters and minimizing connections between them~\cite{stuart2019comprehensive}.

\noindent$\bullet$ The Leiden algorithm is an improved version of Louvain, providing more accurate and stable community detection by addressing the issue of large or unstable clusters and ensuring better quality of partitions~\cite{stuart2019comprehensive}.

\subsection{Deep Learning-based Clustering Models}

\noindent$\bullet$ DEC jointly learns feature representations and cluster assignments via deep neural networks, adopting a clustering-guided loss function to force the generated sample embeddings to have minimal distortion relative to the pre-learned cluster centers~\cite{xie2016unsupervised}.

\noindent$\bullet$ DESC is an unsupervised method that optimizes a clustering objective function through iterative self-learning to simultaneously remove batch effects and reveal biologically meaningful soft clustering structures in scRNA-seq data, without explicitly requiring batch information~\cite{li2020deep}.

\noindent$\bullet$ scDeepCluster, a single-cell model-based deep embedded clustering method, which simultaneously learns feature representation and clustering via explicit modeling of scRNA-seq data generation~\cite{tian2019clustering}.

\noindent$\bullet$ scMAE is a masked autoencoder-based method for scRNA-seq data that perturbs gene expression and uses a masked autoencoder to reconstruct the original data, learning robust and informative cell representations to enhance clustering performance~\cite{fang2024scmae}.

\noindent$\bullet$ scNAME is a scRNA-seq clustering algorithm that integrates a mask estimation task for gene pertinence mining and a neighborhood contrastive learning framework with a global memory bank, enabling robust, scalable, and accurate clustering while enhancing rare cell type detection~\cite{wan2022scname}.

\noindent$\bullet$ scDCC is a principled clustering method for scRNA-seq data that integrates domain knowledge into the clustering process to improve clustering performance and enhance the biological interpretability of cell groups~\cite{tian2021model}.

\noindent$\bullet$ scziDesk is a deep learning-based clustering method for scRNA-seq data that combines a denoising autoencoder with a soft self-training K-means algorithm to learn a cluster-friendly latent space, achieving robust and scalable performance~\cite{chen2020deep}.

\subsection{Graph-based Clustering Models}

\noindent$\bullet$ scGNN is a deep learning framework for scRNA-seq data that models cell–cell relationships with graph neural networks and heterogeneous gene expression patterns with a left-truncated mixture Gaussian model, enhancing gene imputation and cell clustering~\cite{wang2021scgnn}.

\noindent$\bullet$ scDSC is a deep structural clustering method for scRNA-seq data that integrates a ZINB-based autoencoder, a graph neural network module, and a mutual-supervised strategy to learn robust representations and improve clustering accuracy and scalability~\cite{wang2022scdssc}.

\noindent$\bullet$ AttentionAE-sc fuses ZINB-based denoising and graph autoencoder-based topological embeddings through an attention mechanism to learn clustering-friendly cell representations, improving clustering accuracy, stability, and robustness~\cite{li2023attention}.

\noindent$\bullet$ scCDCG integrates soft graph construction with spectral clustering, combining deep cut-informed graph embedding, optimal transport-driven self-supervised learning, and autoencoder-based feature extraction to capture high-order structural information and enhance clustering performance~\cite{xu2024sccdcg}.


\subsection{Biological Foundation Models}

\noindent$\bullet$ scGPT is a generative pre-trained transformer model for single-cell data that captures biological insights by pre-training on over 10 million cells and can be finetuned for tasks like batch integration, cell-type annotation, and gene network inference~\cite{cui2024scgpt}.


\noindent$\bullet$ Geneformer is a pretrained transformer model that enables context-aware gene network inference and disease modeling from limited transcriptomic data~\cite{cui2024geneformer}.

\noindent$\bullet$ GeneCompass is a knowledge-informed cross-species foundation model that integrates prior biological knowledge to uncover gene regulatory mechanisms across species, enhancing understanding and enabling advanced applications in gene regulation and cell fate transition~\cite{yang2024genecompass}.

\section{Extended Experiment Specification}
\label{app:appendix_experiment}
\subsection{Experimental Setup}
To ensure robust results, we test each method on all datasets and repeat the experiments with 5 random seeds per method-dataset combination, collecting both the average score and standard deviation to investigate performance and stability.
The experiments are conducted on a computing infrastructure featuring dual Intel(R) Xeon(R) Platinum 8380 CPUs running at 2.30GHz, with a total of 1TB of memory and four NVIDIA A800-80GB GPUs. The system is powered by Ubuntu 18.04 LTS with CUDA 12.4. 

\subsection{Implementation Details}
\subsubsection{Traditional Clustering Models.} 
We evaluated three representative traditional clustering methods implemented in R: SC3, Leiden, and Louvain. To apply these methods to our single-cell RNA-seq datasets, raw data stored in HDF5 format were converted into appropriate R objects.
Specifically, we created SingleCellExperiment objects for SC3 and Seurat objects for Leiden and Louvain. 
This preprocessing step ensured compatibility with subsequent analysis while accommodating diverse data formats, including sparse matrices.

\noindent$\bullet$ \textbf{\textit{SC3}} clustering was performed using fixed, conservative parameters that disabled gene filtering and biological annotation, allowing consistent treatment across datasets. The number of clusters \(k\) was set individually for each dataset based on prior biological knowledge, ranging from 4 to 39 clusters. For large datasets, we adopted a random subsampling strategy, selecting 8,000 cells per run and repeating the clustering five times to improve robustness.

\noindent$\bullet$ \textbf{\textit{Leiden and Louvain} } clustering shared a common preprocessing workflow implemented in \texttt{Seurat}, which included normalization, identification of variable features, scaling, principal component analysis using the top 20 components, and construction of a nearest neighbor graph. Clustering resolution was fixed at 0.5, with the algorithm parameter distinguishing Leiden ({algorithm = 4) from Louvain (algorithm = 1). To handle large-scale data efficiently, we leveraged sparse matrix representations and implemented result caching to avoid redundant computations.

\subsubsection{Deep learning and Graph-based Clustering Models.} 
For both graph-based and deep learning-based clustering models, we adhere strictly to the parameter configurations reported in their original publications. In cases where parameter settings are not explicitly specified, we conduct preliminary experiments to identify a set of hyperparameters that yield stable and broadly effective performance across all datasets. This ensures that each method is evaluated under conditions that are as close as possible to its optimal performance. 

\noindent$\bullet$ \textbf{\textit{DEC}} is a deep learning model that utilizes greedy layer-wise pretraining. During this process, the weights are initialized from a zero-mean Gaussian distribution with a standard deviation of 0.01, and each layer is pretrained for 50,000 iterations with 2\% dropout. The deep autoencoder is then fine-tuned for 100,000 iterations without dropout, using a minibatch size of 256, a starting learning rate of 0.1 (which is divided by 10 every 20,000 iterations), and no weight decay.

\noindent$\bullet$ \textbf{\textit{scDeepCluster}} is implemented in Python 3 using Keras with a TensorFlow backend, applying Gaussian noise with a noise level of 2.5, and the autoencoder is pre-trained for 400 epochs using AMSGrad Adam (lr=0.001, $\beta_1$=0.9, $\beta_2$=0.999) before clustering with Adadelta (lr=1.0, rho=0.95). The encoder’s hidden layers are set to (256, 64) with a 32-dimensional latent space, a batch size of 256, and a convergence threshold of 0.1\% delta clustering labels per batch. 

\noindent$\bullet$ \textbf{\textit{DESC}} is trained in two stages: pretraining and clustering. In the pretraining phase, the model runs for 300 epochs with a batch size of 256 and a learning rate of 0.01, using the SGD optimizer, while a dropout rate of 0.2 is applied to prevent overfitting. During the clustering stage, early stopping is utilized based on the loss improvement, with adjustable Louvain clustering resolutions (default values of 0.6 and 0.8), and a k-nearest neighbor graph is constructed using 15 neighbors for local connectivity.

\noindent$\bullet$ \textbf{\textit{scziDesk}} is trained in two stages: a pretraining phase of 1000 epochs and a self-training phase of 2000 epochs, both using a batch size of 256. The learning rate is set to 0.0001, with loss weighting parameters $\alpha=0.001$, $\gamma=0.001$, and noise standard deviation fixed at 1.5. The network architecture adopts a four-layer encoder with hidden dimensions of (500, 256, 64, 32), progressively reducing feature dimensionality to learn compact cell representations.

\noindent$\bullet$ {\textit{scDCC}} uses a ZINB autoencoder with hidden layer sizes (256, 64, 32, 64, 256) and a bottleneck of 32, applying Gaussian noise with a standard deviation of 2.5. The model is pretrained for 300 epochs using AMSGrad Adam (lr=0.001, $\beta_1$=0.9, $\beta_2$=0.999), and clustering is optimized with Adadelta (lr=1.0, $\rho$=0.95), with a batch size of 256 and $\gamma$ and constraint loss weight both set to 1. Clustering stops when the proportion of changed labels per epoch falls below 0.1\%. 

\noindent$\bullet$ \textbf{\textit{scNAME}} uses a ZINB autoencoder architecture consisting of layers (256, 64, 32), where 32 is the bottleneck size. Pretraining is conducted for 500 epochs using a corruption probability of 0.3 and Gaussian noise with a standard deviation of 1.5, followed by fine-tuning for 1000 epochs with clustering updates every 50 epochs. The model is trained with a batch size of 128 and learning rate of 0.0001, and uses a combination of three loss components weighted by $\alpha=1.0$, $\beta=0.1$, and $\gamma=0.1$. 

\noindent$\bullet$ \textbf{\textit{scMAE}} uses a masked autoencoder with a hidden size of 128 and a latent dimension of 32. The model is trained for 100 epochs with a batch size of 256 using Adam optimizer (lr=1e-3), where masked input is generated using a corruption probability of 0.4 per gene and a mask loss weight of 0.7. Clustering is performed at epoch 80 using either KMeans or Leiden (depending on sample size), and the number of clusters is fixed to the known number of cell types. 

\noindent$\bullet$ \textbf{\textit{scGNN}} adopts an autoencoder trained in two stages: an initial regularization stage for 500 epochs, followed by 10 EM iterations, each including 200 epochs of training and cluster refinement using KMeans. The model is trained with a batch size of 12,800 and an Adam optimizer (lr=0.001). A KNN graph (k=10) is constructed using Euclidean or cosine distance, and latent representations can be further enhanced via GAE embedding before final clustering. 

\noindent$\bullet$ \textbf{\textit{scDSC}} integrates a pre-trained autoencoder with a GNN-based clustering module. The autoencoder consists of symmetric layers (3276, 1000, 1000, 4000) with three latent layers ($z_1=2000$, $z_2=500$, $z_3=10$), and is pretrained for 100 epochs using Adam (lr=1e-4). After initialization, the model is trained for 80 epochs using RAdam with a composite loss combining reconstruction (MSE), ZINB loss, KL divergence, and binary cross-entropy, with weights (1, 0.1, 0.01, 0.1) respectively; KMeans is used to initialize cluster centers and guide the learning process based on a GNN-enhanced latent representation. 

\noindent$\bullet$ \textbf{\textit{AttentionAE-sc}} combines a multi-head attention-based autoencoder with graph-regularized clustering. The encoder and decoder are symmetric with hidden layers [256, 64, 64, 256], and a latent dimension of 16 is used. The model is trained in two stages: a reconstruction stage for 200 epochs using Adam (lr=0.001), followed by 100 clustering epochs guided by a target distribution and neighborhood graph constructed via Gaussian kernel. 

\noindent$\bullet$ \textbf{\textit{scCDCG}} is trained in two stages, where both the pretraining and clustering phases adopt identical hyperparameter settings, including learning rate and weight decay. The encoder is designed with layers [256,16] to produce a 16-dimensional latent representation, while the decoder follows a symmetric structure, and the network is optimized using Adam for 200 epochs. The smoothness parameter of the optimal transport strategy is fixed at $\lambda=5$, the degrees of freedom of the Student’s t-distribution is set to $\theta=1$, and dataset-specific hyperparameters are determined by grid search.

\subsubsection{Biological Foundation Models.}
We use three foundation models: \textit{scGPT}, pre-trained on 33 million human single-cell transcriptomes; \textit{Geneformer}, pre-trained on 30 million human cells; and \textit{GeneCompass}, pre-trained on 100 million human and mouse cells.

\paragraph{\textit{Foundation Model for Clustering.}}  
Embeddings from the pre-trained models are extracted and subjected to K-means clustering with a predefined number of clusters corresponding to annotated cell types. scGPT generates token-level embeddings from a 4-layer Transformer; Geneformer uses mean pooling from a 6-layer Transformer; and GeneCompass outputs length-normalized embeddings from its 12-layer architecture, with clustering run for up to 300 iterations.

\paragraph{\textit{Foundation Model for Classification.}}  
All models are fine-tuned using the same 8:2 train-validation split to ensure comparability across experiments.

\noindent$\bullet$ \textbf{\textit{scGPT}.} Fine-tuned for 10 epochs with a batch size of 32, learning rate of $1 \times 10^{-4}$, and a dropout rate of 0.2, using a 4-layer Transformer with 128-dimensional hidden embeddings.  

\noindent$\bullet$ \textbf{\textit{Geneformer}.} Fine-tuned for 50 epochs with a batch size of 16, learning rate of $5 \times 10^{-5}$, and linear decay with 100-step warm-up.  

\noindent$\bullet$ \textbf{\textit{GeneCompass}.} Fine-tuned for 30 epochs with a batch size of 1, learning rate of $5 \times 10^{-5}$, and a 100-step warm-up, utilizing dynamic batching and weight decay for regularization.

\subsection{Evaluation Metrics}
\label{app:appendix_evaluation}
The assessment of clustering performance relies on three established metrics from the public domain: Accuracy (ACC), Normalized Mutual Information (NMI)~\cite{strehl2002cluster}, and Adjusted Rand Index (ARI)~\cite{vinh2009information}. 
Higher values of these metrics indicate better clustering performance. 
Given the knowledge of the ground truth class assignments $U$ and our clustering algorithm assignment $V$ on $n$ data points. 
\subsubsection{Accuracy (ACC).} ACC measures the best matching between two cluster assignments $U$ and $V$. Given a data point $i$, with ground truth label $l_i$ and clustering assignment $u_i$, ACC is defined as:
\begin{equation}
    \text{ACC}=\max_{m}\frac{\sum_{i=1}^n 1 {l_i=m(u_i)}}{n},
\label{equ:ca}
\end{equation}
here, $m$ ranges over all possible one-to-one mapping between $U$ and $V$, the best mapping with respect to $m$ can be efficiently found using the Hungarian algorithm~\cite{kuhn1955hungarian}.\par
\subsubsection{Normalized Mutual Information (NMI).} The NMI is a function that measures the consistency between the predicted and true labels of $n$ cells. Specifically,
\begin{equation}
\tiny
    \text{NMI}=\frac{\sum_{i=1}^{C_U} \sum_{j=1}^{C_V} \frac{\mid U_i\bigcap V_j\mid}{n}\text{log}\frac{n\mid U_i\bigcap V_j\mid}{\mid U_i\mid\mid V_j\mid}}{\text{mean}(-\sum_{i=1}^{C_U} \mid U_i\mid \text{log} \frac{\mid U_i\mid}{n},-\sum_{j=1}^{C_V} \mid V_j\mid \text{log} \frac{\mid V_j\mid}{n})}.
\label{equ:nmi}
\end{equation}
\par
\subsubsection{Adjusted Rand Index (ARI).} The ARI is a function that evaluates the similarity of the two assignments ignoring permutations. Given a set $X$ of size $n$ and two clustering results $U = \{u_1, u_2,\cdots, u_r\}$, $V = \{v_1, v_2,\cdots, v_s\}$, we denote $n_{i j}=\left|U_i \cap V_j\right|, i=1,2, \ldots, r, j = 1, 2,\cdots, s$. We also let $a_i=\sum_{i=1}^r n_{i j}$, $b_j=\sum_{j=1}^s n_{i j}$. Then we evaluate:
\begin{equation}
\tiny
\text{ARI} = \frac{\sum_{i,j} \binom{n_{i,j}}{2} - \left[ \sum_i \binom{a_i}{2} \sum_j \binom{b_j}{2} \right] \binom{n}{2}}{\left[ \sum_i \binom{a_i}{2} \sum_j \binom{b_j}{2} \right]/ 2 - \left[ \sum_i \binom{a_i}{2} \sum_j \binom{b_j}{2} \right]/\binom{n}{2}}.
\label{equ:ari}
\end{equation}
\par

\subsubsection{F1 Score (Macro-F1).} 
The macro-averaged F1 score evaluates performance by computing the F1 score for each class independently and then taking the average. Let $P_k$ and $R_k$ denote the precision and recall for class $k$, respectively. Then the macro-F1 score is:
\begin{equation}
\text{F1}_{\text{macro}} = \frac{1}{K} \sum_{k=1}^K \frac{2 \cdot P_k \cdot R_k}{P_k + R_k},
\label{equ:f1}
\end{equation}
where $K$ is the total number of classes. This metric gives equal importance to each class regardless of its size.

\begin{table*}[!th]
\centering
\resizebox{0.99\textwidth}{!}{
\setlength{\tabcolsep}{2pt}
\renewcommand{\arraystretch}{1}
\begin{tabular}{lcccccccccccccccc}
\toprule
\multicolumn{1}{c}{} &
\multicolumn{3}{c}{Traditional} & &
\multicolumn{7}{c}{Deep Learning-based} & &
\multicolumn{4}{c}{Graph-based} \\
\cmidrule{2-4} \cmidrule{6-12} \cmidrule{14-17}

& SC3 & louvain & leiden &
& DEC & DESC & scDeepCluster & scMAE & scNAME & scDCC & scziDesk &
& scGNN & scDSC & AttentionAE-sc & scCDCG \\
 
 \midrule
 
Human Pancreas 1
 & 71.27±0.90 & 79.35±1.21 & 81.16±0.07 & 
 & 61.43±0.07 & 83.30±0.81 & 76.05±3.31 & \underline{86.01±2.72} & 83.29±4.83 & 78.94±3.03 & 68.17±0.42 & 
 & 64.38±0.98 & 56.42±3.74 & 82.87±3.64 & \textbf{86.35±0.94} \\
 
Human Pancreas 2
 & 70.38±1.48 & \textbf{89.87±0.05} & \underline{89.73±1.97} & 
 & 65.96±1.84 & 77.93±2.46 & 73.62±0.14 & 80.27±3.30 & 79.10±2.61 & 77.68±2.44 & 59.10±0.58 & 
 & 65.97±1.59 & 70.75±2.46 & 85.10±5.28 & 83.44±2.55 \\
 
Human Pancreas 3
 & 69.48±2.00 & 88.73±0.10 & \textbf{90.09±0.03} & 
 & 64.92±1.21 & \underline{89.55±2.20} & 74.37±1.20 & 87.54±1.95 & 81.74±1.25 & 77.44±1.42 & 79.45±7.12 & 
 & 63.13±2.17 & 72.57±0.55 & 86.93±2.48 & 82.21±1.06 \\
 
Human Pancreas4
 & 66.45±1.37 & 77.27±0.17 & 77.13±0.32 & 
 & 62.72±3.41 & 79.93±1.91 & 74.13±0.70 & \underline{82.36±3.64} & 79.42±1.80 & 75.72±0.50 & 65.82±7.02 & 
 & 53.52±0.00 & 70.56±2.83 & \textbf{83.17±4.27} & 81.94±0.54 \\
 
Mauro Human Pancreas cells
 & \underline{88.50±0.47} & 84.70±0.00 & \textbf{89.96±0.12} & 
 & 71.96±2.00 & 79.63±1.37 & 78.89±0.98 & 88.49±0.44 & 85.31±5.11 & 83.78±4.70 & 64.94±14.21 & 
 & 63.54±3.16 & 63.00±8.23 & 86.57±2.12 & 86.81±0.98 \\
 
68K PBMC
 & 70.99±0.15 & 65.17±0.11 & 68.48±0.71 & 
 & 61.40±2.80 & 70.21±0.71 & \underline{77.86±2.34} & 75.50±0.20 & 74.97±0.31 & \textbf{78.62±1.09} & 64.66±1.28 & 
 & 36.16±2.80 & 35.30±0.00 & 64.90±1.21 & 73.49±0.67 \\
 
CITE CMBC
 & 66.40±2.34 & \textbf{82.17±1.87} & \underline{78.78±1.65} & 
 & 36.83±12.97 & 72.44±1.77 & 73.41±1.61 & 74.00±2.66 & 67.69±2.19 & 73.82±0.96 & 49.31±14.47 & 
 & 65.58±2.63 & 13.91±0.89 & 66.00±10.86 & 74.77±2.66 \\
 
Human Kidney
 & \underline{79.65±2.12} & 66.62±1.98 & 70.37±1.76 & 
 & 30.82±4.36 & 72.24±2.65 & 65.84±2.26 & \textbf{82.24±1.89} & 76.78±0.76 & 63.06±2.37 & 78.01±2.24 & 
 & 25.20±0.00 & 18.33±0.00 & 69.33±4.72 & 67.98±2.62 \\
 
Sonya liver
 & 79.51±2.23 & 64.08±1.87 & 70.70±1.65 & 
 & 48.54±3.86 & 77.67±0.51 & 79.69±0.68 & \textbf{85.59±1.44} & \underline{85.04±3.65} & 80.78±1.90 & 83.34±0.70 & 
 & 23.03±0.00 & 6.31±0.65 & 83.23±5.31 & 71.34±2.59 \\
 
Sapiens Liver
 & 77.95±1.98 & \textbf{79.98±1.76} & 70.15±2.12 & 
 & 71.92±1.90 & 60.95±0.00 & 60.98±2.20 & \underline{78.25±0.60} & 74.83±0.70 & 70.80±2.00 & 75.10±1.20 & 
 & 77.21±2.00 & 68.21±5.20 & 51.17±8.40 & 62.54±3.20 \\
 
Sapiens Ear Crista Ampullaris
 & 55.95±0.00 & 76.93±0.00 & 73.42±0.00 & 
 & 46.07±6.01 & 61.60±0.00 & 50.39±1.50 & 74.31±0.20 & 74.53±0.10 & \underline{78.26±1.40} & 0.00±0.00 & 
 & \textbf{79.27±1.10} & 72.72±9.50 & 64.98±8.90 & 69.54±2.90 \\
 
Sapiens Ear Utricle
 & 65.58±0.00 & \underline{76.00±0.00} & 71.20±0.00 & 
 & 53.80±2.10 & 28.26±0.00 & 43.87±1.20 & \textbf{78.28±1.20} & 75.56±1.30 & 69.00±2.90 & 74.21±2.30 & 
 & 64.62±4.90 & 74.69±8.60 & 59.40±11.10 & 66.82±5.40 \\
 
Sapiens Lung
 & 67.24±0.00 & 70.82±0.00 & 70.16±0.00 & 
 & 69.67±4.00 & 70.73±0.00 & 54.53±0.70 & \underline{79.65±0.60} & 75.20±0.70 & 73.20±1.10 & 78.38±0.40 & 
 & \textbf{81.31±0.90} & 65.39±3.80 & 57.47±8.50 & 66.94±1.90 \\
 
Sapiens Testis
 & 35.50±0.00 & 44.29±0.00 & 44.26±0.00 & 
 & 47.04±3.00 & 44.12±0.00 & 33.57±3.40 & 57.09±0.40 & \textbf{60.57±3.00} & 55.27±1.60 & 53.60±1.20 & 
 & \underline{59.01±0.50} & 43.70±7.20 & 47.44±3.40 & 57.42±3.30 \\
 
Sapiens Trachea
 & 54.60±0.00 & 65.98±0.00 & 63.81±0.00 & 
 & 63.70±5.00 & 69.03±0.00 & 58.14±0.80 & \underline{77.12±1.50} & \textbf{78.04±1.60} & 71.06±1.30 & 71.04±2.50 & 
 & 73.05±0.00 & 57.42±5.80 & 0.00±4.80 & 63.25±1.00 \\
 
\midrule

Mouse cerebral cortex
 & \underline{71.57±0.00} & 56.05±0.00 & 60.86±0.00 & 
 & 45.14±8.41 & 59.89±0.88 & \textbf{73.44±0.19} & 60.15±0.87 & 64.77±1.00 & 70.64±3.92 & 61.38±3.22 & 
 & 23.53±1.46 & 4.05±1.16 & 66.13±0.70 & 58.25±1.06 \\
 
Mouse embryonic stem
 & \textbf{100.00±0.00} & 76.70±2.34 & 77.87±1.98 & 
 & 65.79±10.09 & 79.36±1.71 & 92.74±0.77 & 72.03±0.57 & 79.95±1.08 & 69.05±11.77 & 77.07±1.13 & 
 & 44.67±1.83 & 54.78±0.00 & 77.90±1.85 & \underline{95.61±0.20} \\
 
Mouse hypothalamus
 & 38.99±3.21 & 0.29±0.12 & 0.23±0.09 & 
 & 33.35±2.81 & 56.41±1.72 & 52.99±0.56 & \textbf{73.79±0.15} & 72.49±2.15 & 53.30±0.85 & \underline{73.79±0.84} & 
 & 22.04±0.00 & 25.71±0.70 & 70.13±3.00 & 64.76±0.61 \\
 
Mouse Pancreas 1
 & 62.70±2.87 & 74.21±1.76 & 76.25±2.12 & 
 & 53.35±2.54 & 76.47±2.83 & 66.46±0.47 & \textbf{82.95±2.54} & 76.44±1.73 & 70.47±0.92 & 75.10±1.73 & 
 & 58.77±1.49 & 27.38±2.51 & \underline{79.92±2.71} & 77.66±2.02 \\
 
Mouse Pancreass 2
 & 52.98±2.45 & 69.87±1.98 & 72.92±1.76 & 
 & 52.98±1.86 & 68.59±1.80 & 63.30±0.93 & \underline{86.66±0.67} & 78.01±1.96 & 69.66±1.83 & 71.55±11.86 & 
 & 55.70±2.61 & 56.72±1.96 & 84.67±4.60 & \textbf{88.02±0.18} \\
 
Shekhar mouse retina
 & 75.53±2.12 & 82.45±1.65 & 84.27±1.43 & 
 & 36.87±3.26 & \underline{88.54±2.67} & 80.73±0.92 & \textbf{91.50±0.05} & 88.43±0.61 & 82.30±0.88 & 31.69±0.49 & 
 & 33.55±1.66 & 8.61±6.68 & \textcolor{gray}{OOM} & 76.71±1.47 \\
 
Macosko mouse retina
 & 68.23±2.34 & 76.23±1.87 & 73.89±2.12 & 
 & 49.96±1.27 & \underline{86.56±0.90} & 77.04±0.72 & \textbf{88.36±0.50} & 84.62±0.21 & 79.73±0.90 & 79.24±1.96 & 
 & 33.50±0.00 & 37.72±0.00 & \textcolor{gray}{OOM} & 68.71±2.52 \\
 
Mouse Kidney
 & \underline{85.67±1.76} & 70.84±2.45 & 77.58±1.98 & 
 & 11.02±2.65 & 78.92±0.00 & 76.19±5.25 & \textbf{86.95±0.33} & 83.07±3.14 & 78.19±0.96 & 85.19±4.98 & 
 & 7.23±0.23 & 4.07±4.04 & 24.10±27.15 & 51.32±2.74 \\
 
Mouse bladder
 & 74.81±2.12 & \underline{78.60±1.65} & 77.28±1.87 & 
 & 62.02±1.04 & \textbf{78.74±1.73} & 73.69±1.72 & 75.57±1.91 & 74.20±1.22 & 75.58±0.58 & 48.96±12.12 & 
 & 55.69±1.26 & 42.24±2.01 & 66.70±3.27 & 75.91±0.93 \\
 
QS Diaphragm
 & 91.39±1.23 & 75.19±2.34 & 75.29±1.98 & 
 & 37.04±3.75 & 80.22±3.06 & 79.91±1.54 & \textbf{95.38±0.36} & 91.94±1.17 & 23.20±11.75 & 61.22±5.40 & 
 & 27.97±3.62 & 15.59±0.93 & 91.80±1.41 & \underline{94.57±0.82} \\
 
QS Lung
 & 63.62±2.45 & 67.06±1.87 & 66.35±2.12 & 
 & 36.52±4.38 & 76.51±0.64 & 71.76±1.76 & \textbf{83.31±0.88} & \underline{78.58±0.49} & 35.81±32.24 & 70.66±3.60 & 
 & 24.11±1.78 & 19.46±4.51 & 77.97±5.77 & 70.13±1.92 \\
 
QS Trachea
 & 55.88±2.34 & 43.74±2.87 & 52.09±2.12 & 
 & 25.77±1.38 & 56.94±1.13 & 58.61±3.09 & 65.58±16.54 & \underline{69.00±6.89} & 28.23±24.98 & 63.20±7.84 & 
 & 10.75±9.36 & 3.96±4.53 & \textbf{74.80±5.02} & 61.91±0.05 \\
 
QS Limb Muscle
 & 86.65±1.76 & 70.33±2.45 & 76.72±1.98 & 
 & 35.98±0.85 & 76.42±3.45 & 76.71±2.61 & \textbf{96.95±2.32} & \underline{94.17±1.01} & \textcolor{gray}{ERR} & 78.57±9.96 & 
 & 26.86±3.93 & 2.08±0.96 & 85.97±4.39 & 83.56±1.37 \\
 
Qx Limb Muscle
 & \textbf{97.80±0.87} & 65.04±2.34 & 71.96±2.12 & 
 & 75.52±1.22 & 77.09±1.68 & 86.69±1.71 & \underline{96.87±2.16} & 96.22±0.68 & 86.34±1.01 & 92.88±1.82 & 
 & 37.75±0.65 & 46.02±0.00 & 94.67±0.90 & 90.24±1.38 \\
 
Qx Bladder
 & 93.56±1.23 & 41.69±2.87 & 44.62±2.45 & 
 & 60.77±1.70 & 54.59±1.06 & 77.52±1.78 & 82.58±11.42 & 91.01±9.45 & 73.02±5.68 & \textbf{97.25±0.47} & 
 & 54.18±3.65 & 44.82±2.40 & 89.03±6.03 & \underline{94.35±1.30} \\
 
Qx Spleen
 & 50.49±2.76 & 35.43±2.34 & 36.69±2.12 & 
 & 36.25±16.03 & 47.50±1.22 & 59.04±1.73 & \textbf{86.00±0.67} & \underline{85.33±4.06} & 58.07±9.82 & 84.97±2.11 & 
 & 7.31±1.62 & 14.99±10.28 & 54.13±46.81 & 74.87±1.38 \\
 
Muris Limb Muscle
 & \textbf{86.38±1.45} & 0.15±0.08 & 0.24±0.12 & 
 & 51.49±7.20 & 64.13±0.00 & 34.55±4.90 & 59.44±3.80 & 63.51±1.30 & \underline{68.79±3.20} & 36.66±6.40 & 
 & 22.10±4.70 & 30.30±4.50 & 12.55±6.10 & 56.54±7.60 \\
 
Muris Brain
 & \textbf{61.12±2.87} & 17.18±2.34 & 26.52±2.65 & 
 & 1.86±1.50 & 7.35±0.00 & \underline{59.48±44.40} & 1.33±0.00 & 0.01±0.00 & 5.52±0.40 & 0.00±0.00 & 
 & 0.31±0.10 & 0.07±0.00 & 2.31±1.01 & 22.48±8.30 \\
 
Muris Kidney
 & 58.51±2.45 & 19.75±2.12 & 21.38±1.98 & 
 & 50.17±2.70 & \textbf{64.80±1.77} & 35.14±5.60 & 54.37±1.90 & 46.94±2.40 & \underline{59.51±2.30} & 39.29±2.70 & 
 & 41.84±1.60 & 21.67±2.90 & 23.76±8.40 & 55.82±1.30 \\
 
Muris Liver
 & 53.10±2.34 & 54.24±2.12 & 50.59±1.87 & 
 & 50.21±2.10 & \underline{64.87±0.00} & 45.07±2.20 & \textbf{65.39±1.00} & 61.21±1.70 & 61.25±3.70 & 43.17±3.40 & 
 & 55.15±1.10 & 35.07±4.70 & 20.59±10.70 & 62.06±2.40 \\
 
Muris Lung
 & 53.60±2.76 & 49.62±2.45 & 64.32±2.12 & 
 & 64.64±2.90 & \textbf{70.81±0.00} & 49.40±1.70 & 64.49±0.90 & 55.95±0.60 & \underline{65.96±1.10} & 12.14±15.00 & 
 & 41.95±2.50 & 23.69±5.40 & 8.88±5.60 & 49.53±3.80 \\

\midrule
 
Average
 & 69.50±3.12 & 62.13±3.87 & 63.81±2.65 & 
 & 49.82±7.37 & 68.12±1.28 & 65.72±5.37 & 76.57±0.64 & 74.66±1.06 & 66.34±1.13 & 62.47±2.49 & 
 & 43.89±1.09 & 36.34±1.28 & 61.61±8.23 & 71.33±4.03 \\

\midrule
 
Rank of average
 & 4 & 10 & 8 & 
 & 12 & 5 & 7 & 1 & 2 & 6 & 9 & 
 & 13 & 14 & 11 & 3\\
 
\bottomrule
\end{tabular}
}
\caption{NMI scores (mean ± std) across 36 datasets. The best results are bolded, and the second-best are \underline{underlined}.}
\label{tab:NMI_AVG±STD_label}
\end{table*}

\begin{table*}[!th]
\centering
\resizebox{0.99\textwidth}{!}{
\setlength{\tabcolsep}{2pt}
\renewcommand{\arraystretch}{1}
\begin{tabular}{lcccccccccccccccc}
\toprule
\multicolumn{1}{c}{Dataset} &
\multicolumn{3}{c}{Traditional} & &
\multicolumn{7}{c}{Deep Learning-based} & &
\multicolumn{4}{c}{Graph-based} \\
\cmidrule{2-4} \cmidrule{6-12} \cmidrule{14-17}

& SC3 & louvain & leiden & 
& DEC & DESC & scDeepCluster & scMAE & scNAME & scDCC & scziDesk &
& scGNN & scDSC & AttentionAE-sc & scCDCG\\
    
\midrule

Human Pancreas 1
& 57.91±2.25 & 68.74±2.42 & 66.95±0.18 &
& 24.71±0.23 & 59.30±0.69 & 48.68±4.64 & \underline{85.23±6.38} & 67.02±17.19 & 52.21±7.51 & 42.55±0.59 &
& 38.54±0.90 & 45.61±2.21 & 73.67±10.70 & \textbf{92.83±0.56} \\

Human Pancreas 2
& 61.45±2.50 & \textbf{92.88±0.08} & \underline{92.73±4.00} &
& 39.89±1.44 & 61.85±5.01 & 50.15±1.25 & 80.14±5.97 & 63.77±6.50 & 63.86±4.36 & 32.65±1.09 &
& 49.04±4.59 & 73.86±1.38 & 79.90±7.57 & 85.30±8.55 \\

Human Pancreas 3
& 64.84±3.22 & \underline{91.71±0.27} & \textbf{96.51±0.08} &
& 44.31±2.06 & 88.69±6.09 & 49.26±1.78 & 90.27±4.44 & 68.83±5.73 & 59.60±6.98 & 72.59±18.56 &
& 58.72±1.45 & 79.78±3.07 & 89.80±3.24 & 90.78±0.65 \\

Human Pancreas4
& 55.44±2.12 & 69.99±0.53 & 69.96±1.00 &
& 38.06±5.81 & 63.47±4.54 & 50.48±1.55 & \underline{80.12±8.10} & 61.00±4.41 & 49.16±3.34 & 51.91±20.29 &
& 27.08±0.00 & 70.60±7.36 & 79.80±6.95 & \textbf{83.62±0.70} \\

Mauro Pancreas
& \underline{93.02±0.82} & 90.71±0.00 & \textbf{93.57±0.12} &
& 55.02±2.58 & 66.44±2.31 & 64.90±1.07 & 92.38±0.35 & 84.51±13.42 & 82.06±12.77 & 50.31±29.92 &
& 50.22±1.91 & 57.48±9.07 & 88.90±4.65 & 91.37±1.21 \\

68K PBMC
& 72.15±0.55 & 56.98±0.30 & 57.52±0.44 &
& 46.69±2.67 & 57.96±0.96 & 72.59±8.12 & 68.04±0.04 & 71.20±2.99 & \textbf{75.59±2.23} & 45.38±2.90 &
& 21.13±2.69 & 17.39±0.00 & 43.60±2.51 & \underline{73.07±1.42} \\

CITE CMBC
& 48.09±2.94 & \underline{62.76±4.48} & \textbf{66.56±3.90} &
& 20.55±12.95 & 55.07±1.70 & 51.49±1.73 & 60.06±5.44 & 52.98±2.84 & 56.62±1.95 & 23.87±20.06 &
& 50.43±3.61 & 3.91±0.70 & 49.37±12.91 & 61.46±1.42 \\

Human Kidney
& \underline{73.27±3.57} & 59.65±0.01 & 62.68±2.35 &
& 22.30±2.95 & 53.66±5.67 & 50.49±5.44 & \textbf{75.57±4.63} & 62.58±0.51 & 46.35±5.40 & 69.03±2.69 &
& 19.00±0.00 & 8.92±0.00 & 50.67±4.81 & 64.91±0.88 \\

Sonya liver
& 75.08±2.66 & 43.85±0.05 & 54.86±0.53 &
& 23.84±4.40 & 49.31±1.13 & 69.71±1.18 & \textbf{88.92±1.78} & \underline{83.96±13.63} & 68.28±2.53 & 83.87±1.17 &
& (0.44)±0.00 & 1.48±0.06 & 77.23±15.55 & 81.26±2.69 \\

Sapiens Liver
& 58.71±10.56 & 60.27±0.20 & 48.05±0.00 &
& 65.47±4.10 & 33.65±0.00 & 34.52±2.00 & 63.82±4.10 & 58.94±2.50 & 48.05±6.90 & \underline{66.12±1.90} &
& \textbf{72.36±1.70} & 56.94±7.90 & 42.41±12.70 & 41.11±4.40 \\

Sapiens Ear Crista Ampullaris
& \textbf{78.83±11.44} & 75.34±0.00 & 59.38±0.31 &
& 41.31±7.83 & 28.38±0.00 & 37.21±2.30 & 57.35±0.10 & 57.60±0.20 & 73.28±4.20 & \textcolor{gray}{ERR} &
& \underline{75.74±1.40} & 63.68±14.30 & 60.47±11.20 & 66.16±4.80 \\

Sapiens Ear Utricle
& 44.59±0.00 & 63.36±0.00 & 62.76±0.00 &
& 42.04±4.90 & 26.12±0.00 & 28.05±1.00 & \underline{64.09±0.70} & 63.46±1.80 & 56.61±5.60 & 63.92±0.70 &
& 56.20±6.10 & \textbf{75.79±10.30} & 46.33±12.60 & 60.16±3.30 \\

Sapiens Lung
& 45.94±0.00 & 45.06±1.36 & 44.19±0.37 &
& 54.51±8.50 & 47.57±0.00 & 30.13±1.40 & 58.40±1.60 & 51.43±2.40 & 48.89±2.80 & \underline{67.58±1.20} &
& \textbf{75.46±2.60} & 54.84±11.60 & 42.16±7.10 & 60.15±1.50 \\

Sapiens Testis
& 30.22±0.00 & 51.59±0.10 & 51.52±0.00 &
& 30.73±2.90 & 14.77±0.00 & 18.91±3.10 & 43.27±0.50 & \underline{55.52±10.20} & 38.79±0.80 & 51.22±1.20 &
& \textbf{58.81±0.90} & 47.20±12.50 & 52.16±3.60 & 55.38±7.20 \\

Sapiens Trachea
& 32.10±0.00 & 52.31±0.47 & 41.80±2.34 &
& 41.14±14.00 & 30.79±0.00 & 27.67±0.00 & 53.69±4.80 & \underline{68.71±6.70} & 39.62±4.50 & 62.58±10.60 &
& 58.10±0.00 & 60.93±9.30 & \textbf{86.31±3.96} & 42.92±2.40 \\

\midrule

Mouse cerebral cortex
& \textbf{78.89±0.39} & 51.18±1.49 & 57.11±0.00 &
& 34.83±13.72 & 42.93±2.92 & 65.18±0.24 & 56.57±0.69 & \underline{66.13±1.29} & 62.62±7.30 & 58.76±6.53 &
& 10.28±0.92 & 0.82±0.08 & 59.30±2.59 & 54.56±0.46 \\

Mouse embryonic stem
& \textbf{100.00±0.06} & 83.22±0.00 & 83.52±0.00 &
& 55.98±11.44 & 77.47±2.03 & 94.94±0.60 & 71.41±1.10 & 80.09±0.67 & 59.67±16.55 & 79.53±1.89 &
& 31.01±0.88 & 47.94±0.00 & 78.07±4.36 & \underline{96.46±1.55} \\

Mouse hypothalamus
& 45.74±1.08 & (0.64)±0.02 & (0.28)±0.03 &
& 34.22±5.02 & 33.25±2.87 & 45.28±1.19 & \textbf{77.61±0.86} & 74.64±4.95 & 51.14±3.30 & \underline{76.05±0.83} &
& 13.11±0.00 & 13.99±0.74 & 72.17±7.67 & 69.79±0.98 \\

Mouse Pancreas 1
& 46.69±1.63 & 66.10±7.94 & 73.03±0.21 &
& 30.64±3.84 & 55.62±6.86 & 34.95±1.47 & \underline{75.83±10.54} & 52.20±5.40 & 43.02±4.93 & 60.80±2.32 &
& 36.33±4.90 & 14.56±4.50 & 69.94±4.52 & \textbf{80.59±5.78} \\

Mouse Pancreass 2
& 31.78±2.37 & 56.29±0.01 & 58.31±2.44 &
& 17.79±2.11 & 36.67±2.32 & 33.35±2.22 & \underline{90.55±0.56} & 61.76±5.86 & 43.87±1.95 & 57.97±27.73 &
& 26.17±3.42 & 55.54±2.62 & 86.07±6.12 & \textbf{92.61±0.28} \\

Shekhar mouse retina
& 64.41±1.79 & 70.34±0.00 & 70.41±1.31 &
& 16.52±2.91 & 84.65±15.63 & 53.95±4.03 & \textbf{95.92±0.07} & \underline{94.28±0.38} & 59.22±3.85 & 13.47±0.27 &
& 20.86±1.65 & 2.76±4.90 & \textcolor{gray}{OOM} & 60.65±6.47 \\

Macosko mouse retina
& 42.93±2.91 & 80.74±2.74 & 67.12±0.05 &
& 26.08±3.18 & \textbf{89.61±0.86} & 42.19±2.58 & \underline{88.95±1.00} & 82.99±2.33 & 55.21±9.32 & 68.38±2.55 &
& 17.49±0.00 & 32.17±0.00 & \textcolor{gray}{OOM} & 58.99±2.69 \\

Mouse Kidney
& \textbf{87.16±1.27} & 66.59±0.00 & 78.79±0.71 &
& 5.65±1.61 & 68.09±0.00 & 67.73±4.25 & \underline{86.85±0.40} & 79.64±6.52 & 70.58±1.43 & 82.98±9.91 &
& 2.10±0.18 & 1.09±1.02 & 10.57±16.08 & 36.85±2.19 \\

Mouse bladder
& \underline{70.56±6.48} & 65.94±0.06 & \textbf{80.53±1.31} &
& 39.33±1.71 & 66.26±5.35 & 54.29±5.20 & 53.83±9.07 & 52.55±3.67 & 56.29±2.10 & 20.62±16.43 &
& 37.94±1.56 & 23.69±2.43 & 44.43±4.46 & 64.55±2.32 \\

QS Diaphragm
& 97.12±0.06 & 68.72±0.20 & 68.62±0.06 &
& 21.79±4.13 & 62.81±6.95 & 65.21±1.09 & \textbf{97.46±0.28} & 95.33±1.53 & 16.34±9.16 & 44.91±14.78 &
& 18.00±2.23 & 6.97±0.95 & 95.33±1.40 & \underline{97.20±0.46} \\

QS Lung
& 48.48±0.50 & 50.34±0.00 & 48.67±0.00 &
& 20.31±3.74 & 48.30±1.01 & 43.69±4.61 & 76.28±0.58 & 67.36±1.88 & 29.79±27.43 & 64.78±5.51 &
& 16.51±1.06 & 9.05±1.83 & \underline{79.53±6.92} & \textbf{80.67±0.78} \\

QS Trachea
& 51.13±3.60 & 25.52±0.00 & 37.52±2.03 &
& 17.91±1.96 & 22.59±2.53 & 49.94±2.34 & 60.41±21.83 & 64.34±14.06 & 25.93±22.50 & \underline{70.62±9.32} &
& 11.07±8.53 & 0.44±0.49 & 68.90±14.70 & \textbf{76.79±0.07} \\

QS Limb Muscle
& 94.51±15.95 & 59.55±0.00 & 66.52±0.04 &
& 26.60±0.65 & 50.07±8.48 & 61.77±5.05 & \textbf{97.74±0.79} & \underline{96.78±0.63} & \textcolor{gray}{ERR} & 77.60±15.72 &
& 15.41±1.95 & 0.30±0.36 & 84.10±11.62 & 90.04±1.20 \\

Qx Limb Muscle
& \textbf{98.98±6.23} & 50.69±0.00 & 64.39±0.00 &
& 70.63±0.79 & 51.03±5.48 & 78.14±2.89 & 97.37±2.83 & \underline{98.27±0.43} & 84.34±3.93 & 95.32±1.86 &
& 33.52±1.25 & 35.82±0.00 & 97.40±0.66 & 93.99±1.93 \\

Qx Bladder
& \textbf{98.76±12.49} & 32.92±0.01 & 37.52±0.01 &
& 62.25±3.31 & 25.25±1.58 & 74.29±0.45 & 80.52±15.01 & 90.66±13.13 & 68.40±5.21 & \underline{98.54±0.40} &
& 56.46±2.12 & 43.12±4.38 & 90.93±6.39 & 98.23±0.79 \\

Qx Spleen
& 51.38±3.54 & 24.76±0.01 & 24.98±0.01 &
& 21.59±9.40 & 18.07±1.70 & 44.15±1.36 & \textbf{92.58±0.53} & 92.22±1.39 & 47.24±18.02 & \underline{92.30±1.01} &
& 8.92±3.15 & 20.64±14.89 & 57.80±50.77 & 86.49±1.57 \\

Muris Limb Muscle
& 50.70±0.00 & 8.75±0.27 & 8.54±0.66 &
& 35.94±10.20 & 33.53±0.00 & 35.93±8.10 & 51.54±3.30 & \underline{54.70±2.30} & \textbf{63.20±4.50} & 32.59±7.20 &
& 21.90±3.20 & 34.28±6.40 & 13.63±6.70 & 53.37±8.50 \\

Muris Brain
& \textbf{93.85±0.00} & (1.07)±0.06 & (1.46)±0.01 &
& 0.51±0.60 & 0.82±0.00 & \underline{60.11±46.80} & (2.22)±0.00 & 0.34±0.10 & 3.14±0.60 & \textcolor{gray}{OOM} &
& 2.77±0.50 & (0.37)±0.80 & 1.97±1.69 & 35.56±7.80 \\

Muris Kidney
& \textbf{46.86±0.96} & 6.60±0.01 & 18.43±1.44 &
& 37.15±3.20 & \underline{44.50±4.21} & 19.47±4.80 & 35.79±1.40 & 32.24±2.90 & 42.29±4.90 & 25.85±5.30 &
& 29.78±0.80 & 14.37±3.20 & 14.66±8.00 & 42.88±2.10 \\

Muris Liver
& 35.29±2.54 & \textbf{48.35±0.22} & 38.63±0.09 &
& 33.18±5.50 & 35.09±0.00 & 27.92±3.90 & \underline{47.55±0.60} & 38.55±3.40 & 37.50±3.70 & 27.27±4.60 &
& 45.58±3.20 & 43.74±12.30 & 3.95±15.10 & 46.96±3.70 \\

Muris Lung
& 22.54±0.00 & 22.16±2.43 & 31.22±5.55 &
& 34.46±4.30 & \textbf{37.20±0.00} & 20.89±1.80 & \underline{35.69±2.40} & 21.79±1.60 & 30.68±1.90 & 2.89±10.00 &
& 32.93±6.50 & 14.56±6.30 & 3.81±3.10 & 26.46±3.80 \\

\midrule
Average
& 62.48±2.96 & 53.40±0.71 & 55.03±0.88 &
& 34.28±4.74 & 47.80±2.75 & 48.82±3.93 & 70.27±3.41 & 65.79±4.60 & 51.70±6.23 & 56.91±7.14 &
& 33.29±2.11 & 31.50±4.39 & 58.69±8.61 & 69.28±2.70 \\

\midrule

Rank of average
& 4 & 8 & 7 &
& 12 & 11 & 10 & 1 & 3 & 9 & 6 &
& 13 & 14 & 5 & 2\\
    
\bottomrule
\end{tabular}
}
\caption{ARI scores (mean ± std) across 36 datasets. The best results are bolded, and the second-best are \underline{underlined}.}
\label{tab:ARI_AVG±STD_label}
\end{table*}

\begin{table*}[!thbp]
\centering
\resizebox{0.95\textwidth}{!}{
\begin{tabular}{@{}ll*{8}{c}@{}}
\toprule
\multirow{2}{*}{Method Type} & \multirow{2}{*}{Method} & \multicolumn{3}{c}{Clustering Quality} & \multicolumn{2}{c}{Performance} & \multicolumn{2}{c}{Stability} & \multirow{2}{*}{Overall} \\
\cmidrule(lr){3-5} \cmidrule(lr){6-7} \cmidrule(lr){8-9}
 & & ACC & NMI & ARI & Train. & Infer. & Seed & Repr. & \\
\midrule
\multirow{3}{*}{Traditional}
 & SC3 & $\times$ & $\times$ & $\times$ & $\checkmark$ & $\checkmark$ & $\times$ & $\times$ & $\times$ \\
 & Louvain & $\times$ & $\times$ & $\times$ & $\checkmark$ & $\checkmark$ & $\times$ & $\times$ & $\times$ \\
 & Leiden & $\times$ & $\times$ & $\times$ & $\checkmark$ & $\checkmark$ & $\times$ & $\times$ & $\times$ \\
\midrule
\multirow{7}{*}{Deep Learning-based}
 & DEC & $\times$ & $\times$ & $\times$ & $\times$ & $\times$ & $\times$ & $\times$ & $\times$ \\
 & DESC & $\times$ & $\times$ & $\times$ & $\times$ & $\times$ & $\times$ & $\times$ & $\times$ \\
 & scDeepCluster & $\checkmark$ & $\checkmark$ & $\checkmark$ & $\times$ & $\checkmark$ & $\times$ & $\times$ & $\times$ \\
 & scMAE & $\checkmark$ & $\checkmark$ & $\checkmark$ & $\checkmark$ & $\checkmark$ & $\checkmark$ & $\checkmark$ & $\checkmark$ \\
 & scNAME & $\times$ & $\times$ & $\times$ & $\times$ & $\checkmark$ & $\times$ & $\times$ & $\times$ \\
 & scDCC & $\times$ & $\times$ & $\times$ & $\times$ & $\checkmark$ & $\times$ & $\times$ & $\times$ \\
 & scziDesk & $\times$ & $\times$ & $\times$ & $\times$ & $\checkmark$ & $\times$ & $\times$ & $\times$ \\
\midrule
\multirow{4}{*}{Graph-based}
 & scGNN & $\times$ & $\times$ & $\times$ & $\times$ & $\times$ & $\times$ & $\times$ & $\times$ \\
 & scDSC & $\checkmark$ & $\checkmark$ & $\checkmark$ & $\times$ & $\checkmark$ & $\times$ & $\times$ & $\times$ \\
 & AttentionAE-sc & $\times$ & $\times$ & $\times$ & $\times$ & $\checkmark$ & $\times$ & $\times$ & $\times$ \\
 & scCDCG & $\checkmark$ & $\checkmark$ & $\checkmark$ & $\checkmark$ & $\checkmark$ & $\checkmark$ & $\checkmark$ & $\checkmark$ \\
\bottomrule
\end{tabular}
}
\caption{Multi-dimensional Evaluation of Single-Cell Clustering Methods (Extended Results).}
\label{tab:strict_evaluation}
\end{table*}

\section{Extended Observation and Analysis}
\label{app:appendix_entend observation}
Full results for cell cluster visualization, cell representation distinguishability, marker-overlap calculation, and tracksplot for top-10 marker genes.

\subsection{Quantitative Analysis}

\label{app:appendix_overall performance}
Here, we provide an extended evaluation of clustering performance by incorporating additional experimental results for NMI, and ARI metrics. Specifically, we report the mean and standard deviation (mean ± std) of these scores across all datasets, offering a comprehensive assessment of clustering accuracy and label consistency. The detailed results are summarized in Tab.~\ref{tab:NMI_AVG±STD_label}, and \ref{tab:ARI_AVG±STD_label}. These supplementary evaluations offer a more detailed and rigorous validation of model performance, complementing the primary findings reported in the main text. 

\subsection{Qualitative Analysis}

We present the full results for qualitative analysis in Fig.~\ref{fig:bar_all} and Fig.~\ref{fig:vis_all}. From these results we can the quality of learned embeddings are consistent across most datasets for each model. From Fig.~\ref{fig:vis_all} we discover that DESC frequently generate inconsistent number of clusters compared with the ground-truth annotation. Although DESC uses automatic number of clusters, making these results reasonable, other models with automatic number of clusters, namely DEC and AttentionAE-sc yield clusters more consistent with the ground-truth annotations. Such mismatch influences the performance of DESC, which ranks the 12th in terms of ACC. However, this provides significant room for improvement through marker-overlap based annotation. As shown in Tab.~\ref{tab:result_correction}, DESC's gain from correction is 28.1\% points, ranking the top among all models.

\subsection{Biological Analysis}

\subsubsection{Marker Gene Identification.}

Taking the clustering results of the scCDCG model on the Mauro Pancreas as a representative example, we conducted an in-depth biological analysis to elucidate the interpretability of the clustering outcomes. 
To this end, we further provided tracksplots of the top 10 ranked marker genes, facilitating a more comprehensive examination of expression patterns across clusters. 
As discussed in the main text, clusters 1 and 8 exhibit highly similar expression profiles among their top-expressed genes, wherein the top 3 marker genes alone are insufficient to distinctly separate these two clusters. 
Extended visualization of the top 10 marker genes reveals substantial overlap in expression patterns between clusters 1 and 8, suggesting that these clusters likely correspond to subtypes within a broader cell lineage rather than representing entirely distinct cellular populations.

\subsubsection{Marker-overlap Calculation.}

We present the visualized marker-overlap calculation results of all models on all datasets in Fig.~\ref{fig:overlap_all}. 
We observe that DESC frequently assigns cell to more clusters than the ground-truth assignation, especially for \textit{Muris brain}, \textit{Muris Limb Muscle}, \textit{Sapiens Ear Utricle}, and \textit{Sapiens Testis}. AttentionAE-sc yields the most (3) perfect results on \textit{Muris Brain}, \textit{Sapiens Ear Crista Amoyllaris}, and \textit{Sapiens Testis}.

\subsubsection{Annotation Comparison.} 

To further elucidate differences between cell type annotation strategies, we constructed Sankey diagrams for all models on the \textit{Mauro Pancreas} (Fig.~\ref{fig:sky}). 
These diagrams visualize the assignment flow of cell types, mapping the transitions from model-predicted clusters to their corresponding annotations generated by the best-mapping and marker-overlap strategies, and subsequently, to the gold-standard reference labels. By tracking these data flows, we can systematically analyze how different models exhibit annotation discrepancies under distinct strategies, thereby revealing inconsistencies and potential misclassifications relative to the ground truth cell types.

\section{Extended Evaluation Protocol}
\label{sec:evaluation_framework}

We extend the evaluation framework introduced in the main text to provide a more comprehensive assessment of single-cell clustering methods. This enhanced protocol systematically examines model performance across three key aspects: clustering quality, computational efficiency, and algorithmic stability.
Tab.~\ref{tab:strict_evaluation} presents the expanded evaluation metrics and detailed analyses, supplementing the primary benchmarking results reported earlier.

\subsection{Clustering Quality Assessment}
\label{subsec:clustering_quality}

We employ a multi-metric evaluation strategy to assess clustering performance comprehensively using three established metrics: Accuracy (ACC), Normalized Mutual Information (NMI), and Adjusted Rand Index (ARI).
Methods must satisfy thresholds across all metrics to receive approval ($\checkmark$):
\begin{itemize}
    \item ACC must achieve $\geq 75\%$ in at least 80\% of the 36 evaluated datasets;
    \item ACC must achieve $\geq 75\%$ in at least 80\% of the 36 evaluated datasets;
    \item ARI demands $\geq 55\%$ with an average ranking $\leq 5$ across all datasets.
\end{itemize}
These thresholds ensure that only methods demonstrating robust performance across diverse biological contexts receive positive evaluation.

Among the 14 evaluated methods, only 4 methods (28.6\%) successfully meet these requirements: scMAE, scDeepCluster, scDSC, and scCDCG. Traditional clustering methods (SC3, Louvain, Leiden) consistently fail across all three metrics, while deep learning-based methods show mixed performance. Graph-based methods demonstrate the highest success rate within their category, with both scDSC and scCDCG achieving approval.

\subsection{Computational Performance Evaluation}
\label{subsec:computational_performance}

Our computational assessment encompasses both training efficiency and inference scalability. 
\begin{itemize}
    \item Training efficiency standards are method-specific: deep learning methods must converge within 200 training epochs, while traditional clustering methods must complete processing within 25 minutes for standard datasets. Additional considerations include memory scalability, GPU accessibility, and hyperparameter sensitivity. 
    \item Inference scalability requires linear or sub-linear time complexity with respect to cell count, batch processing capability for large-scale datasets, distributed computing support, and real-time processing capabilities.
\end{itemize}

Training efficiency shows a clear performance divide, with traditional methods (SC3, Louvain, Leiden) universally achieving approval due to algorithmic simplicity. 
Among advanced methods, only scMAE and scCDCG satisfy requirements, resulting in 35.7\% overall success rate (5/14 methods). Inference scalability demonstrates better performance with 64.3\% approval rate (9/14 methods), including all traditional methods and several advanced approaches. 
The performance gap reveals the fundamental trade-off between algorithmic sophistication and computational feasibility.

\subsection{Algorithmic Stability Assessment}
\label{subsec:algorithmic_stability}

Stability evaluation focuses on reproducibility and robustness through two key metrics: seed robustness and reproducibility. 
\begin{itemize}
    \item Seed robustness requires standard deviation $< 5\%$ across at least 5 independent runs with different random seeds to ensure clustering results are not heavily dependent on initialization.
    \item Reproducibility demands consistent results within 2\% variation across different computational environments, including different hardware configurations and software versions.
\end{itemize}
These criteria ensure reliable deployment in diverse research settings and reproducible scientific findings.

Algorithmic stability represents the most challenging evaluation dimension, with only 2 of 14 methods (14.3\%) achieving approval: scMAE and scCDCG. 
Traditional methods demonstrate poor stability performance, with Louvain clustering showing particularly high variability (ACC standard deviation reaching 16.76\%). 
Most deep learning methods suffer from initialization sensitivity, with only scMAE achieving required stability. Among graph-based methods, only scCDCG demonstrates sufficient stability. This widespread failure highlights a critical limitation where algorithmic sophistication often compromises result reliability.

\subsection{Overall Performance Integration}
\label{subsec:overall_integration}

The overall performance assessment requires simultaneous satisfaction of all evaluation criteria across the three dimensions above. 
A method receives overall approval ($\checkmark$) only when achieving approval in all individual components: all three clustering quality metrics (ACC, NMI, ARI), both computational performance aspects (training efficiency and inference scalability), and both stability requirements (seed robustness and reproducibility). 
This conjunctive evaluation ensures that no critical weakness can be compensated by excellence in other dimensions, setting the most stringent standard for method deployment in single-cell analysis.

The overall evaluation reveals exceptional rarity of comprehensive excellence, with only scMAE and scCDCG achieving approval, representing 14.3\% overall success rate. scMAE demonstrates superior performance through its masked autoencoder architecture that balances clustering quality, computational efficiency, and algorithmic stability. scCDCG achieves comprehensive approval through its robust graph-based approach. The remaining 12 methods fail due to various limitations: traditional methods lack clustering quality despite computational advantages, most deep learning methods suffer from computational or stability issues, and several graph-based approaches fail stability requirements. This establishes scMAE and scCDCG as the sole methods suitable for rigorous single-cell clustering applications.

Tab.~\ref{tab:strict_evaluation} systematically presents the evaluation results for all 14 single-cell clustering methods across the four critical dimensions discussed above.

\section{Extended Related Work}
\label{app:appendix_related work}
\subsection{scRNA-seq Clustering Methods}
Clustering methods for scRNA-seq have evolved from traditional models grounded in low-dimensional distance metrics to techniques leveraging deep learning and graph-based modeling, and most recently, to biological foundation models built upon Transformer architectures. 

Traditional methods such as SC3~\cite{kiselev2017sc3}, Louvain~\cite{stuart2019comprehensive}, Leiden~\cite{stuart2019comprehensive}, pcaReduce~\cite{vzurauskiene2016pcareduce}, and SUSSC~\cite{wang2021suscc} primarily employ consensus clustering or modularity optimization to enhance stability and accuracy. However, their reliance on low-dimensional assumptions and unsupervised learning frameworks limits their capacity to capture intricate cellular heterogeneity.

Deep learning-based methods, particularly those leveraging self-supervised learning, including DEC~\cite{xie2016unsupervised}, DESC~\cite{li2020deep}, scDeepCluster~\cite{tian2019clustering}, scDCC~\cite{tian2021model}, scziDesk~\cite{chen2020deep}, scMAE~\cite{fang2024scmae}, and scNAME~\cite{wan2022scname}, have significantly improved robustness to data sparsity and noise by learning feature representations. However, these methods often overlook intrinsic intercellular structural relationships, leading to instability and limited interpretability in complex clustering scenarios.

Graph-based clustering methods, such as scGNN~\cite{wang2021scgnn}, scDSC~\cite{gan2022deep}, AttentionAE-sc~\cite{li2023attention}, scSiameseClu~\cite{xu2025scsiameseclu}, scCDCG~\cite{xu2024sccdcg}, and scSGC~\cite{xu2025soft}, explicitly incorporate intercellular relationships to enhance clustering accuracy and structural awareness. However, these approaches remain sensitive to graph construction strategies and face computational efficiency challenges.

Recently, the emergence of biological foundation models~\cite{zhang-etal-2025-survey-foundation}, such as scGPT~\cite{cui2024scgpt}, scFoundation~\cite{hao2024large}, GeneFormer~\cite{theodoris2023transfer}, and GeneCompass~\cite{yang2024genecompass}, has opened new avenues for scRNA-seq clustering by leveraging large-scale pretraining to achieve broad generalization across datasets and species. These models exhibit advanced capabilities in gene network inference and cross-species knowledge transfer. However, their clustering performance remains constrained by non-specific task designs, as current architectures primarily focus on general-purpose representation learning rather than clustering-oriented optimization.

\subsection{Benchmarks}
Although various studies have benchmarked scRNA-seq clustering methods, these evaluations often focus on specific methodological categories or isolated analytical aspects, such as parameter sensitivity, cell number estimation, batch effect correction, or spatial transcriptomics. For example,~\cite{krzak2019benchmark} conducted an early systematic assessment of 13 R-based clustering algorithms on real and simulated datasets, highlighting sensitivities to parameter tuning and variations in accuracy and efficiency. Similarly,~\cite{yu2022benchmarking} evaluated the stability of traditional clustering methods across diverse datasets and proposed scCCESS, a stability-based approach for cell type estimation. 
Several studies have further extended benchmarking efforts to specialized domains within single-cell analysis. For instance, spatial clustering methods were systematically compared by~\cite{yuan2024benchmarking}, emphasizing spatial continuity and clustering accuracy, while~\cite{dai2022scimc} benchmarked data imputation techniques, identifying the superior performance of deep learning methods and developing the scIMC platform for broader application. 
Furthermore, the study by~\cite{tran2020benchmark} systematically evaluated the efficiency and correction performance of various batch-effect correction methods, ultimately recommending solutions like Harmony for superior results.

To date, a comprehensive benchmarking framework spanning the full spectrum of clustering approaches, from traditional models to biological foundation models, has yet to be established. 
A unified platform integrating diverse clustering algorithms and evaluation metrics is essential for systematic benchmarking and further methodological advancement in single-cell analysis. 



\section{Code and Data Availability}
The scCluBench benchmarking framework and its associated data and code have been publicly released to facilitate and standardize future research in single-cell clustering.
Our comprehensive repository, available at \url{https://github.com/XPgogogo/scCluBench}, provides the core toolkit, curated datasets, detailed usage manuals, and an extended version of this manuscript. To further support comparative studies, Table~\ref{tab:code_available} lists the original repositories of the baseline methods evaluated in this study, which span a spectrum of paradigms from traditional algorithms to deep learning and biological foundation models.

\begin{table}[!th]
\centering
\scriptsize
\begin{tabular}{ll}
\toprule
\textbf{Model} & \textbf{Code Repository Link} \\
\midrule
Leiden/Louvain & \url{https://github.com/satijalab/seurat} \\
SC3 & \url{https://github.com/hemberg-lab/SC3} \\
DEC & \url{https://github.com/XifengGuo/DEC-keras} \\
DESC & \url{https://eleozzr.github.io/desc} \\
scDeepCluster & \url{https://github.com/ttgump/scDeepCluster} \\
scziDesk & \url{https://github.com/xuebaliang/scziDesk} \\
scDCC & \url{https://github.com/ttgump/scDCC} \\
scNAME & \url{https://github.com/aster-ww/scNAME} \\
scMAE & \url{https://zenodo.org/records/10465991} \\
scGNN & \url{https://github.com/juexinwang/scGNN} \\
scDSC & \url{https://github.com/DHUDBlab/scDSC} \\
AttentionAE-sc & \url{https://github.com/LiShenghao813/AttentionAE-sc} \\
scCDCG & \url{https://github.com/XPgogogo/scCDCG} \\
scGPT & \url{https://github.com/bowang-lab/scGPT} \\
GeneFormer & \url{https://github.com/jkobject/geneformer} \\
GeneCompass & \url{https://github.com/xCompass-AI/geneCompass} \\
\bottomrule
\end{tabular}
\vspace{-3mm}
\caption{Official code repositories of baseline clustering methods included in scCluBench.}
\label{tab:code_available}
\vspace{-5mm}
\end{table}


\begin{figure*}[!th]
    \centering
    \includegraphics[width=1\linewidth]{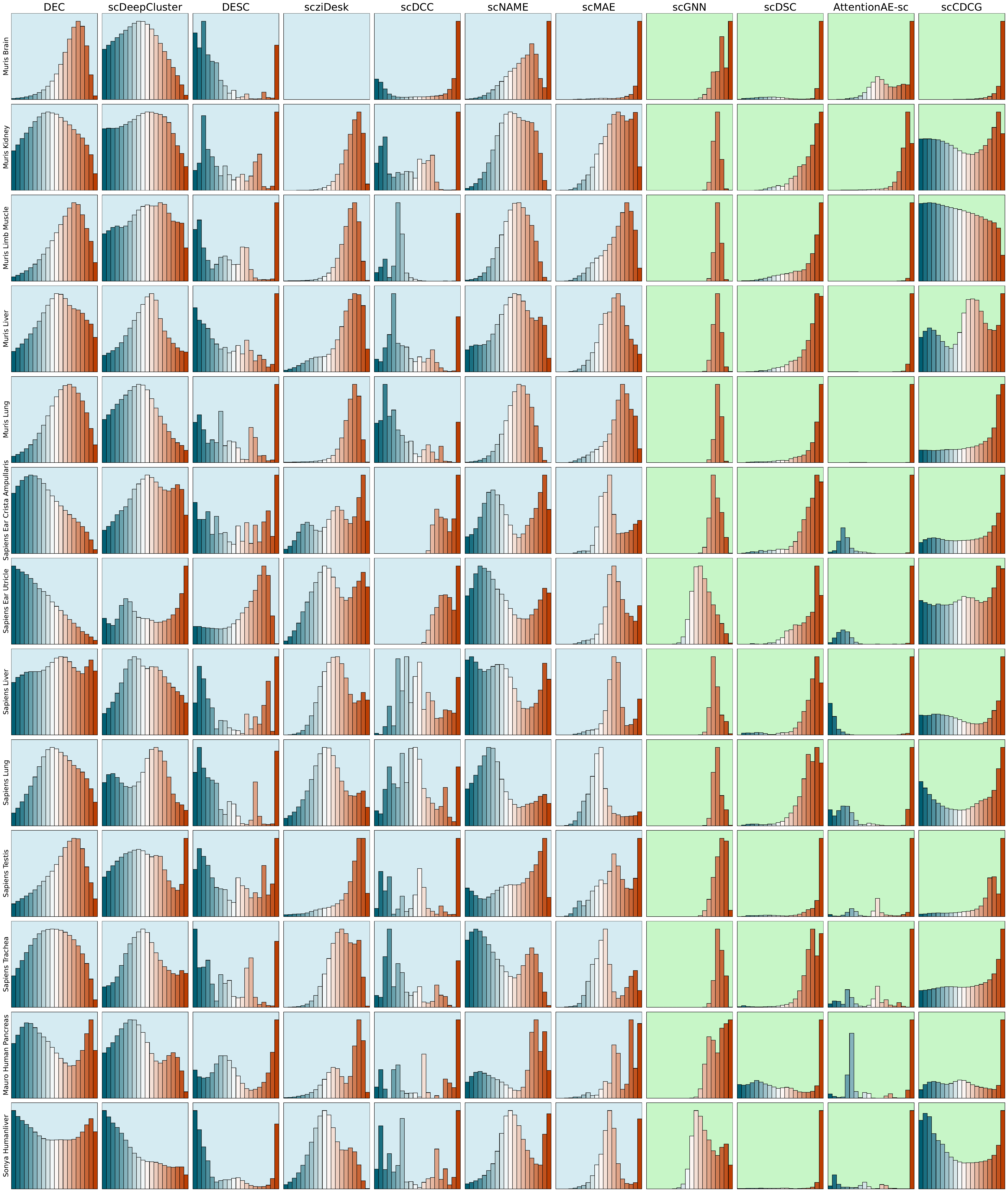}
    \caption{Representation similarity of all models on all datasets.}
    \label{fig:bar_all}
\end{figure*}

\begin{figure*}
    \centering
    \includegraphics[width=1\linewidth]{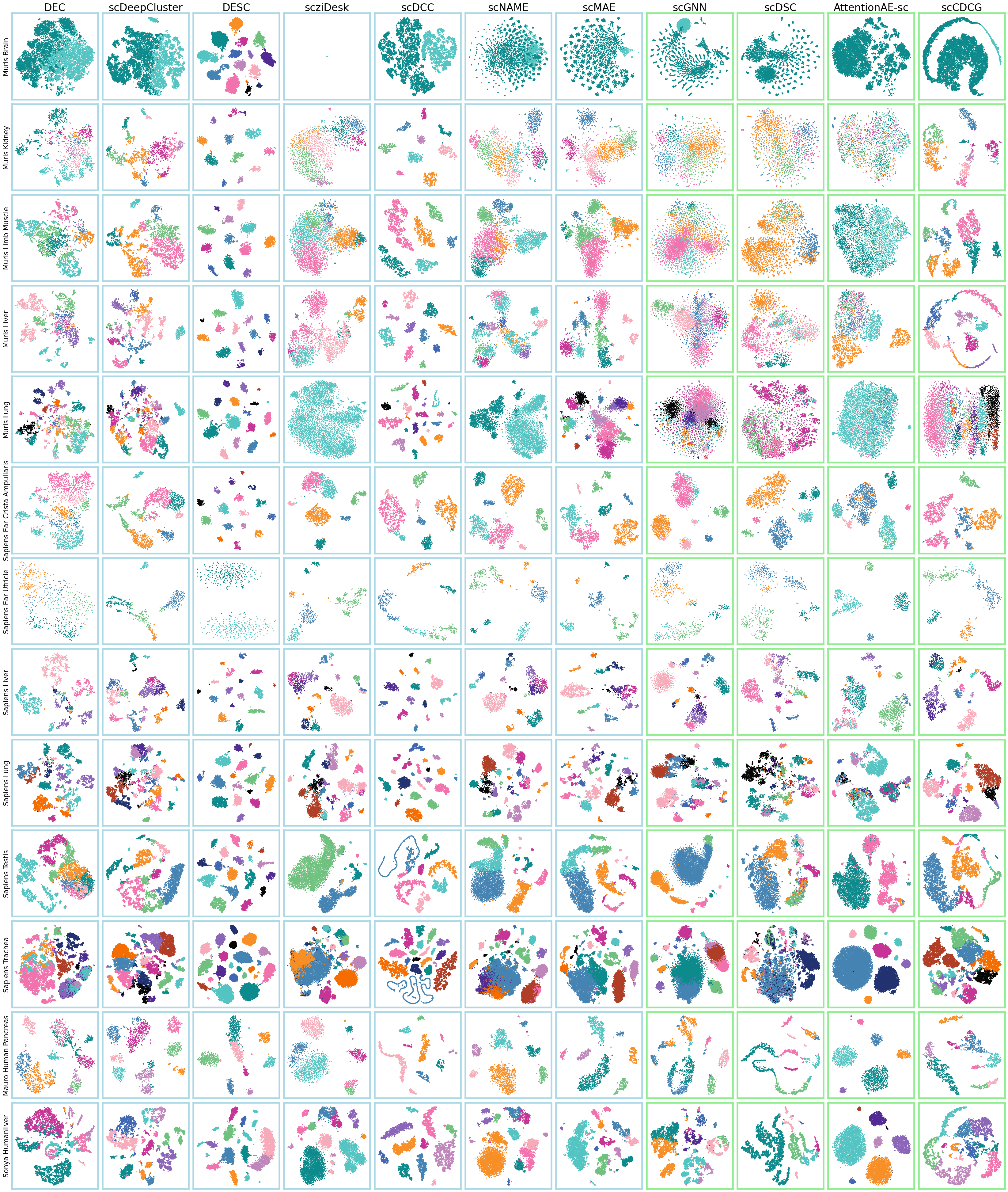}
    \caption{Visualization of all models on all datasets.}
    \label{fig:vis_all}
\end{figure*}

\begin{figure*}
    \centering
    \includegraphics[width=0.65\linewidth]{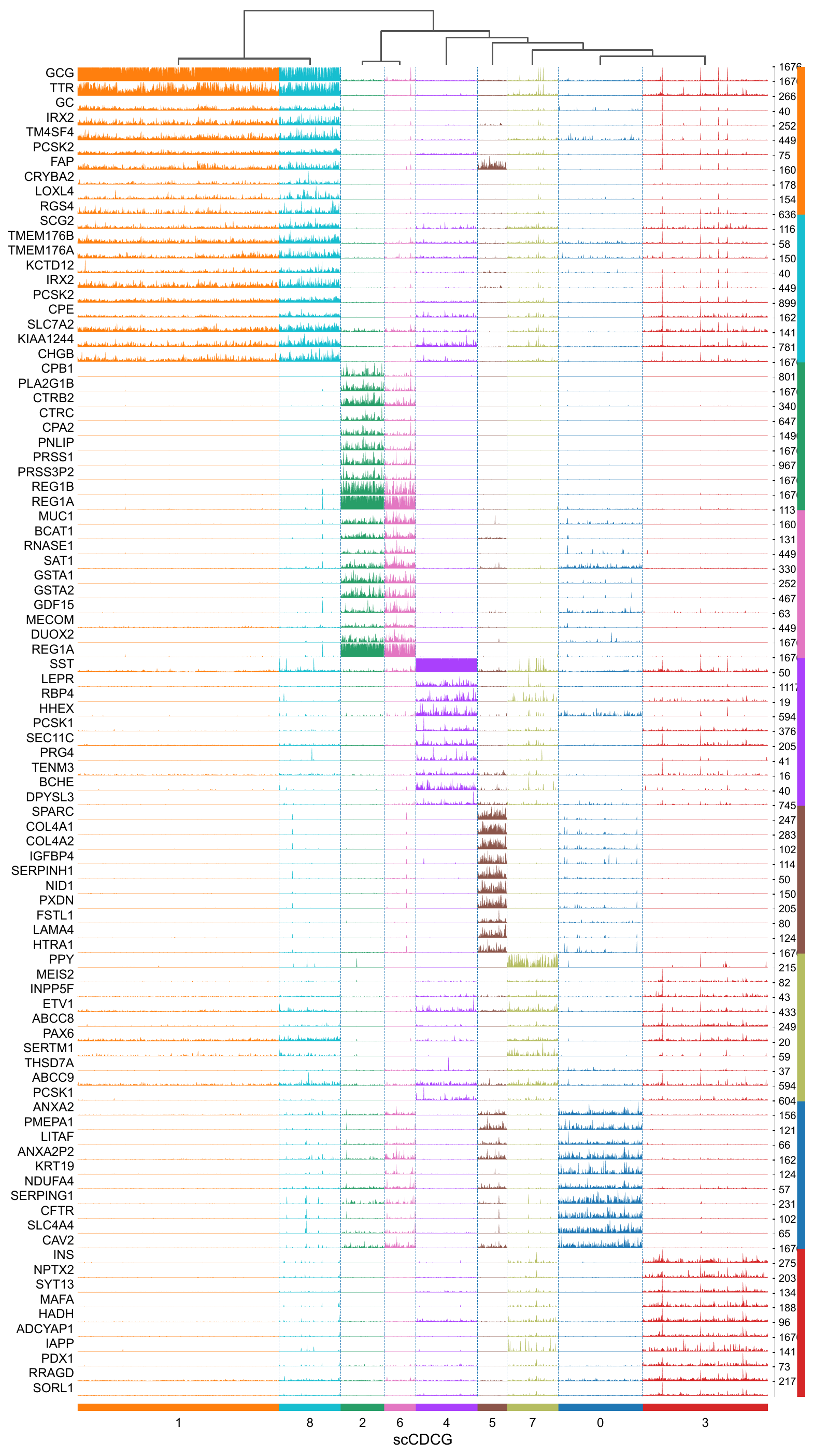}
    \caption{Tracksplot of top 10 marker genes by scCDCG on \textit{Mauro Pancreas}.}
    \label{fig:tracksplotmaurohumanpanceascellsccdcg}
\end{figure*}

\begin{figure*}[htbp]
    \centering
    \includegraphics[width=1\linewidth]{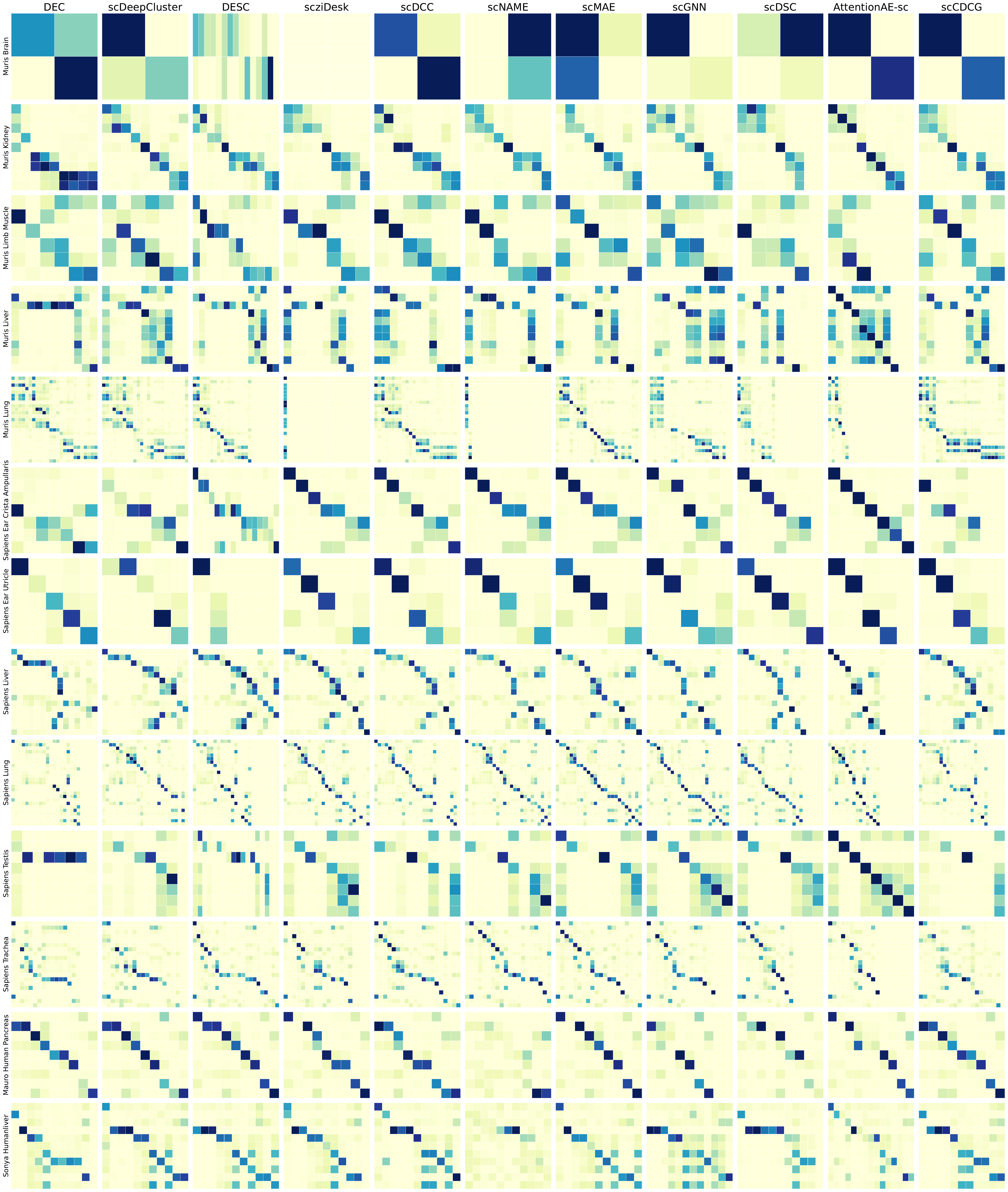}
    \caption{Visualized marker-overlap of all models on all datasets. The X-axis denotes assigned cell type ids by models and the y-axis denote the ground-truth cell type names. Deeper color indicated higher similarity. }
    \label{fig:overlap_all}
\end{figure*}

\begin{figure*}[htbp]
    \centering
    \includegraphics[width=0.75\linewidth]{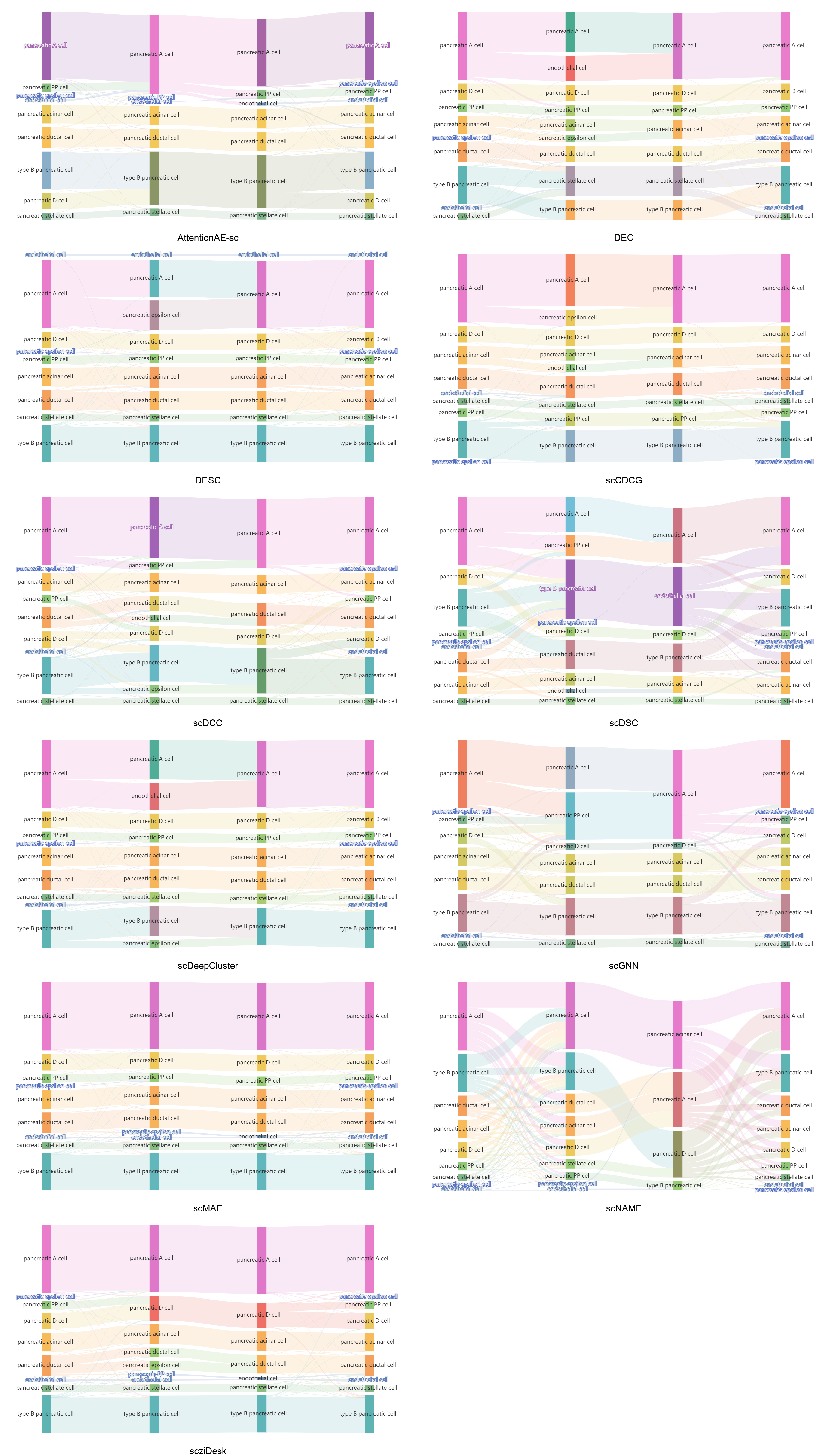}
    \caption{Sankey diagram of all models on \textit{Mauro Pancreas}.}
    \label{fig:sky}
\end{figure*}

\clearpage

\section*{Acknowledgements}
 This work is partially supported by the National Natural Science Foundation of China (Grant No. 92470204 and 62406306), the National Key Research and Development Program of China (Grant No. 2024YFF0729201).

\bibliography{reference}




\end{document}